\documentclass[a4paper,11pt]{article}
\hoffset= - 2cm
\usepackage{amssymb,float,amsmath,amsfonts,amsthm,cases,cite,bm,latexsym,mathrsfs}
\usepackage{eufrak}
\usepackage[normalem]{ulem}
\usepackage{graphicx,color,epsfig}
\usepackage{wrapfig}
\usepackage[pdfencoding=auto]{hyperref}
\usepackage{xcolor}
\usepackage{hyperref}
\definecolor{linkcolor}{HTML}{799B03}
\definecolor{urlcolor}{HTML}{799B03}
\hypersetup{pdfstartview=FitH,linkcolor=linkcolor,urlcolor=urlcolor,colorlinks=true}
\newcommand*{\D}{{\rm d}}

\newcommand*{\tpsi}{\tilde{\psi}}
\newcommand*{\tH}{\tilde{H_2}}
\newcommand*{\bpsi}{\bar{\psi}}
\newcommand*{\bH}{\bar{H_2}}
\newcommand*{\cH}{\mathcal{H}}
\newcommand*{\cG}{\mathcal{G}}
\newcommand*{\cF}{\mathcal{F}}
\newcommand*{\TT}{\mathbf{\Theta}}
\def\[{\begin{equation}}
\def\]{\end{equation}}

\begin{document}

\begin{center}
  {\LARGE \bf  In hot pursuit of a stable wormhole in beyond Horndeski theory}

\vspace{10pt}

\vspace{20pt}
S. Mironov$^{a,b,d}$\footnote{sa.mironov\_1@physics.msu.ru},
\fbox{V. Rubakov$^{a,c}$},
V. Volkova$^{a}$\footnote{volkova.viktoriya@physics.msu.ru}
\renewcommand*{\thefootnote}{\arabic{footnote}}
\vspace{15pt}

% $^a$\textit{Institute for Nuclear Research of the Russian Academy of Sciences,\\
% 60th October Anniversary Prospect, 7a, 117312 Moscow, Russia}\\
% \vspace{5pt}

% $^b$\textit{Moscow Institute of Physics and Technology,\\
% Institutski pereulok, 9, 141701, Dolgoprudny, Russia}

% $^c$\textit{Institute for Theoretical and Mathematical Physics,\\
% M.V. Lomonosov Moscow State University, 119991 Moscow, Russia}

% $^d$\textit{Kurchatov Institute, ITEP\\
% pl. Akademika Kurchatova, 1, 123182 Moscow, Russia}

% $^e$\textit{Department of Particle Physics and Cosmology, Physics Faculty,\\
% M.V. Lomonosov Moscow State University,\\
% Vorobjevy Gory, 119991 Moscow, Russia}
$^a$\textit{Institute for Nuclear Research of the Russian Academy of Sciences, 117312 Moscow, Russia}

$^b$\textit{Institute for Theoretical and Mathematical Physics, MSU, 119991 Moscow, Russia}

$^c$\textit{Department of Particle Physics and Cosmology, Physics Faculty, MSU, 119991 Moscow, Russia}

$^d$\textit{NRC "Kurchatov Institute", 123182, Moscow, Russia}
\end{center}

\vspace{5pt}

\begin{abstract}
We consider the issue of stability at the linearized level for static, spherically symmetric wormhole solutions within a subclass of scalar-tensor theories of beyond Horndeski type. 
In this class of theories we derive 
a set of stability conditions ensuring the absence of ghosts and both radial and angular gradient instabilities about a static, spherically-symmetric background. This set of constraints extends the existing one and completes the stability analysis for high energy modes in both parity odd and parity even sectors, while "slow" tachyonic instabilities remain unconstrained. We give an example of beyond Horndeski Lagrangian admitting a wormhole solution which complies with all stability constraints for the high energy modes.

\end{abstract}

\section{Introduction}
\label{sec:intro}
Wormholes are intriguing gravitational objects whose hypothetical existence has been capturing scientists' attention for decades~\cite{Ellis, Bronnikov,Morris:1988cz,Morris:1988tu,Visser}. 
However, at a classical level having such tunnels in space-time within
General Relativity (GR) is tricky: it was proved necessary
to fill the region near a 
wormhole's throat with a matter which would 
violate the Null Energy Condition (NEC) and prevent 
the throat from collapsing~\cite{Hochberg_Visser}.
It is known that
heathy violation of NEC is quite challenging since
all conventional matter comply with it 
and NEC-violating theories often suffer from instabilities~\cite{Dubovsky:2005xd,RubakovNEC}.

One possible way out, which works for microscopic wormholes, 
is to make use of
quantum effects and, for example, consider charged massless fermions which give rise to a negative Casimir-like 
energy supporting the throat of a wormhole~\cite{Gao:2016bin, Fu:2019vco, Maldacena2018}. 
Another approach 
suitable at a classical level as well
is to directly introduce an exotic NEC-violating matter~\cite{Bronnikov:2013coa}, for instance,
a scalar field with a wrong sign kinetic term 
(a phantom scalar field)~\cite{Kodama,Armendariz,Gonzalez:2008wd}. The latter example of phantom fields is widely used in existing models but turns out unstable in quantum regime~\cite{Cline:2003gs}.

An alternative approach to preventing the wormhole throat from collapsing is to modify GR by adding a 
non-minimally coupled scalar field
and make use of an effective stress-energy tensor arising in such theories to violate NEC 
\footnote{In fact, once gravity is modified NEC is no longer relevant and one replaces it with Null Convergence Condition (NCC)~\cite{Tipler}, which is also hard to violate.}
in a healthy way, 
like e.g. in
$f(R)-$theories~\cite{Lobo}, Brans-Dicke theories~\cite{Agnese,Nandi},
dilaton Einstein-Gauss-Bonnet theories~\cite{Kanti}, etc. 
In fact it's been realised quite recently that 
all viable scalar-tensor theories of modified gravity 
are specific cases of Horndeski theories~\cite{Horndeski:1974wa} and their generalisations, namely, beyond Horndeski 
(or GLPV)~\cite{Gleyzes:2014dya,Gleyzes:2014qga} and DHOST theories~\cite{Langlois1,Langlois2} (see e.g. Ref.~\cite{KobaRev} for review). Horndeski theories and their generalisations  are the most general scalar-tensor theories describing $2+1$ degrees of freedom (DOF) while admitting the second derivatives in the Lagrangian.
These theories proved to be effective for NEC violation with no 
instabilities invoked, so
(beyond) Horndeski theories became quite popular as a framework for constructing spherically-symmetric solutions like wormholes~\cite{Bronnikov:2010tt,Korolev,Rubakov:2015gza,Rubakov:2016zah,Kolevatov:2016ppi,Trincherini,wormhole1,KorolevLobo,Bakopoulos}.

However, there is a subtle issue concerning stability against small perturbations for any spherically-symmetric solution in a scalar-tensor theory of modified gravity. 
The stability
problem for static, spherically-symmetric wormholes in Horndeski theory is formulated as a no-go theorem~\cite{Olegi}, which shows that any 
wormhole solution inevitably 
involves ghost DOF at some point, typically 
close to the throat region.
But the situation considerably changes as soon as one goes beyond Horndeski~\cite{Trincherini,wormhole1}: in generalised Horndeski theories 
the structure of
stability conditions is significantly modified so that the no-go theorem no longer holds. Thus, beyond Horndeski theories, at least in principle, admit static, spherically-symmetric solutions which are free from ghosts throughout the space. In particular, Ref.~\cite{wormhole1} has put forward a specific Lagrangian of beyond Horndeski type, which admits a static wormhole solution that is free from ghosts and radial gradient instabilities at all points. The suggested wormhole, however, cannot claim complete stability, since the derived set of stability constraints did not address gradient instabilities, propagating along the angular direction 
as well as "slow" tachyonic modes in the parity even sector (symmetric under reflection of 2-sphere). 

In this paper we aim to 
extend the existing set of stability conditions of Ref.~\cite{wormhole1}
and derive a constraint eliminating angular gradient instabilities in the even-parity sector.
The latter completes the stability analysis for high energy perturbation modes in both parity odd and parity even sectors. However,
due to technical complexity we still put aside the potential problem of tachyonic instabilities
in the even-parity sector, which despite being significant only for low energy modes might cause troubles 
over long periods of time.

As this work is sequential to Ref.~\cite{wormhole1} it aims to 
address and modify the weak spots of the previously suggested partially stable wormhole solution. 
In particular, in this paper we 
briefly discuss possible gauge choices in the linearized theory in the context of stability analysis for a wormhole. 
Gauge fixing strategy in Ref.~\cite{wormhole1} was inspired 
by previous works on stability constraints in general Horndeski theory
in Refs.\cite{Kobayashi:odd,Kobayashi:even}, whose main motivation concerned black holes' stability. However, similar gauge choice
turned out to be unsuitable for analysing wormhole's stability close to the throat region since right at the narrowest point one of the gauge functions hits infinity. Surely, there is no such problem in this 
gauge for a black hole solution. Thus, in order to make 
the stability analysis conclusive even in the center of a wormhole's throat in this paper we adopt a more appropriate gauge choice for even-parity modes, namely, Regge-Wheeler-unitary gauge, which was used e.g. in Ref.~\cite{Trincherini} but within ADM EFT approach.
In result we rederive the existing stability constraints in a covariant form and get the new ones for angular gradient instabilities.
% In fact, the gauge fixing in the even-parity sector of Ref.~\cite{wormhole1} is unsuitable for analysing wormhole's stability close to the throat region since right at the narrowest point one of the gauge functions hits infinity. 
% The latter makes the gauge choice of Ref.~\cite{wormhole1} not quite conclusive for stability analysis of a wormhole solution.
% Thus, in this paper we adopt a more appropriate gauge choice for even-parity modes (similar to that in Ref.~\cite{Trincherini}), rederive the existing stability constraints in a covariant form and get the new ones for angular gradient instabilities.

Another issue noted already in Ref.~\cite{wormhole1}
but not properly discussed there concerns potential singularity of coefficients in the quadratic action for even-parity modes with high angular momenta. 
% Moreover, by switching the gauge choice we also address the issue 
% of potentially vanishing denominators in some coefficients in the quadratic action for even-parity modes with high angular momenta 
% discovered in Ref.~\cite{wormhole1}. 
%
% These denominators involve the same factor $\TT$ which can  
% take zero values at some points, where corresponding coefficients in the quadratic action hit infinity. But at the same time the dispersion relations for even-parity modes remain perfectly finite when $\TT$ crosses zero, which indicates....
%
It turned out that some coefficients in the quadratic action for two even-parity DOFs 
involve the same factor $\TT$ in denominators,
which inevitably   
takes zero values at some point(s) if no-ghost stability condition is satisfied everywhere. However,
despite seeming singularities in the coefficients  
the dispersion relations for even-parity perturbations remain perfectly finite for any values of $\TT$ including zero. 
The latter indicates that even though there are potential singularities
in the linearized equations the corresponding solutions are not necessarily pathological.
This situation with seeming divergencies upon $\TT$ crossing zero 
has somewhat imperfect analogue in a cosmological setting -- the so called $\gamma$-crossing, discussed in the context of bouncing models in (beyond) Horndeski theory~\cite{Ijjas,GammaCrossing}.
It was explicitly shown in Ref.~\cite{GammaCrossing} that even though
the linearized equations in the unitary gauge (with vanishing scalar field perturbations) involved seeming divergencies of coefficients at $\gamma$-crossing, the corresponding solutions 
for perturbations were perfectly 
regular for any value of the denominator $\gamma$ including zero.
The latter naively indicates that "$\TT$-crossing" in our static, spherically symmetric case also is not related to any pathologies in the theory.
In this paper  
we show explicitly that in fact zero $\TT$ is not a special point and
all perturbations in the even-parity sector behave regularly for any 
values of $\TT$ including zero. 
% Unlike the initially adopted gauge 
% of Ref.~\cite{wormhole1}, $\TT$-crossing manifests itself in Regge-Wheeler-unitary gauge

Finally, in Sec.~\ref{sec:wormhole_example} 
we suggest a specific example of beyond Horndeski Lagrangian which admits a wormhole solution that is stable against high energy perturbations, i.e. ghosts and gradient instabilities propagating in both radial and angular directions. This model is to a significant extent inspired by our previous model in Ref.~\cite{wormhole1} with the main difference of excluding angular gradient instabilities in the even-parity sector of the new model.
Our new wormhole solution is still imperfect: the wormhole 
has a vanishing mass and even though asymptotically far away the space-time corresponds to an empty Minkowski type, the gravity remains modified as compared to GR even at large distances from the wormhole.

The paper is organised as follows. In Sec.~\ref{sec:setup} 
we introduce both the Lagrangian and the the background setting.
We develop the linearized theory for perturbations in Sec.~\ref{sec:linearized_th}, separating parity odd and parity even sectors. 
Sec.~\ref{sec:odd_sector} contains a recap of the existing results for odd-parity perturbations. 
We briefly discuss possible gauge choices for parity even modes 
in the beginning of Sec.~\ref{sec:even_parity} in the context of choosing the most appropriate one for the
wormhole stability analysis. In Sec.~\ref{sec:old_gauge} we revisit
the stability analysis in the initial gauge of Ref.~\cite{wormhole1}
to illustrate the issue with $\TT$-crossing. Then 
Sec.~\ref{sec:new_gauge} presents the original calculations of the quadratic action for perturbations carried out in Regge-Wheeler-unitary gauge. In Sec.~\ref{sec:TT_crossing} we prove that 
the linearized equations in Regge-Wheeler-unitary gauge admit regular
solutions at and around $\TT-$crossing.
We formulate the new set of stability conditions for high energy modes 
in the even-parity sector in Sec.~\ref{sec:new_stability_constraints}
and finally give a specific example of beyond Horndeski Lagrangian which admits a wormhole solution obeying the new set of stability constraints in Sec.~\ref{sec:wormhole_example}.
We briefly conclude in Sec.~\ref{sec:outlook}.

%%%%%%%%%%%%%%%%%%%%%%%%%%%%%%%%%%%%%%%%%%%%%%%%%%%%%%%%%%%%%%%%%%%%%%%%%
\section{Wormholes in beyond Horndeski theories: a background set-up}
\label{sec:setup}

In this section we specify the Lagrangian of the theory and the background setting along with introducing our notations.

We consider a quadratic subclass of beyond Horndeski theories with the following action:
\begin{multline}
\label{eq:lagrangian}
% \mathcal{L}(F,G_4,F_4) 
S
= \int \D^4x \sqrt{-g} \left( F(\pi,X)
% + K(\pi,X)\Box\pi
+ G_4(\pi,X)R \right.\\\left.
+ \left(2 G_{4X}(\pi,X) - 2 F_4 (\pi,X) \; X\right)\left[\left(\Box\pi\right)^2-\pi_{;\mu\nu}\pi^{;\mu\nu}\right] \right.\\\left.
- 2 F_4 (\pi,X) \left[\pi^{,\mu} \pi_{;\mu\nu} \pi^{,\nu}\Box\pi -  \pi^{,\mu} \pi_{;\mu\lambda} \pi^{;\nu\lambda}\pi_{,\nu} \right]\right),
\end{multline}
where $\pi$ is the scalar field,
$X=-\frac12 g^{\mu\nu}\pi_{,\mu}\pi_{,\nu}$,
$\pi_{,\mu}=\partial_\mu\pi$,
$\pi_{;\mu\nu}=\triangledown_\nu\triangledown_\mu\pi$,
$\Box\pi = g^{\mu\nu}\triangledown_\nu\triangledown_\mu\pi$,
$G_{4X}=\partial G_4/\partial X$. The functions $F$ and $G_4$
are characteristic of Horndeski theories, while the
non-vanishing $F_4$ extends the theory to beyond Horndeski
type. 

In what follows we concentrate on a static, spherically-symmetric background with the following metric
\[
\label{eq:backgr_metric}
ds^2 = - A(r)\:dt^2 + \frac{dr^2}{B(r)} + J^2(r) \left(d\theta^2 + \sin^2\theta\: d\varphi^2\right),
\]
where the radial coordinate $r$ runs from $-\infty$ to $+\infty$,
and the functions $A(r)$, $B(r)$ and $J(r)$ are positive and bounded
from below,
\[
\label{eq:metric_bounded}
A(r) \geq A_{min} > 0, \quad B(r) \geq B_{min} > 0, \quad J(r) \geq R_{min} > 0,
\]
with $R_{min}$ standing for the radius of the wormhole throat.
Note that $r$ is a globally defined coordinate, so that the
horizon absence and  ``flaring-out''
conditions are satisfied automatically. 

In what follows we aim to construct an asymptotically flat wormhole 
on both ends, which is the case provided that
\[
A(r) \to 1 \; , \;\;\;\; B(r) \to 1 \; ,
\;\;\;\; J(r) = |r| + O(1) \;\;\;\; \mbox{as} \;\;\; r \to \pm \infty \; .
\]
We choose a specific form of the metric functions 
$A(r)$, $B(r)$ and $J(r)$ which
describe a wormhole in Sec.~\ref{sec:wormhole_example}.
Even though $B(r)$  in eq.~\eqref{eq:backgr_metric}
can be set equal to 1 by coordinate transformation,
we keep it arbitrary for generality.
In our setting the scalar field $\pi$ is static and
depends on the radial coordinate only, $\pi = \pi(r)$ and
hence $X = -B(r) \pi'^2 /2$, where prime denotes the
derivative with respect to $r$.

In what follows we make use of background equations of motion following
from action~\eqref{eq:lagrangian}, especially 
when we reconstruct
the specific form of Lagrangian functions of beyond Horndeski theory which admits a wormhole solution in Sec.~\ref{sec:wormhole_example}. We keep the notations used in Ref.~\cite{wormhole1} for the equations of motion:
\[
\label{eq: background_eqs}
  {\cal E}_A = 0, \quad {\cal E}_B = 0, \quad {\cal E}_J = 0,
  \quad {\cal E}_{\pi} = 0 \; ,
  \]
  where $ {\cal E}_A$ is obtained by varying the action with respect to
  $A$, etc. We give the explicit forms of
${\cal E}_A, {\cal E}_B, {\cal E}_J$ and ${\cal E}_{\pi}$
in Appendix A. 

In the following section we develop the linearized theory aiming to formulate a complete set of stability conditions for high energy modes
about 
a static, spherically symmetric background~\eqref{eq:backgr_metric}.

%%%%%%%%%%%%%%%%%%%%%%%%%%%%%%%%%%%%%%%%%%%%%%%%%%%%%%%%%%%%%%%%%%%%%%%%%
\section{Linearized theory}
\label{sec:linearized_th}
As it was stated above one of our priorities is to ensure stability of a wormhole solution against small perturbations. In what follows we
consider the
linearized metric
\begin{equation}
\label{eq:metric}
g_{\mu\nu} = \bar{g}_{\mu\nu} + h_{\mu\nu},
\end{equation}
where the background $\bar{g}_{\mu\nu}$ is given 
in eq.~\eqref{eq:backgr_metric} and $h_{\mu\nu}$ stand for small metric perturbations, while $\delta\pi$ denotes perturbation about a background scalar field $\bar{\pi}$. 

To study the behaviour of perturbations about a static and spherically-symmetric background we follow a standard approach and make use of the Regge-Wheeler formalism~\cite{ReggeWheeler}:
perturbations are classified into odd-parity and even-parity sectors 
based on their behaviour under two-dimensional reflection.
With further expansion of perturbations into series of the spherical harmonics 
$Y_{\ell m}(\theta,\phi)$ the modes with
different parity, $\ell$ and $m$ do not mix at the linear level, so 
it is legitimate to consider them separately.
% that calculations get considerably simplified. 
% Moreover, due to the symmetry under rotations it is possible to consider only modes with $m=0$ without loss of generality. 
In the next two subsections
we consider odd-parity and even-parity sectors one by one. 

%%%%%%%%%%%%%%%%%%%%%%%%%%%%%%%%%%%%%%%%%%%%%%%%%%%%%%%%%%%%%%%%%%%%%%%
\subsection{Odd-parity sector}
\label{sec:odd_sector}
Even though ghost-like instabilities typically arise in the even-parity 
sector, we give a brief review of the stability constraints in the odd-parity sector for completeness. The contents of this subsection
are a short summary of Sec.3.2 in Ref.~\cite{wormhole1}.   

The scalar field $\pi$ does not obtain odd-parity perturbations, so the only source of perturbation modes in this sector is metric $g_{\mu\nu}$.
The most general form of metric 
perturbations in the odd-parity sector read~\cite{ReggeWheeler}:
\[
\label{odd_parity}
\mbox{Parity~odd}\quad \begin{cases}
\begin{aligned}
 & h_{tt}=0,~~~h_{tr}=0,~~~h_{rr}=0,\\
 & h_{ta}=\sum_{\ell, m}h_{0,\ell m}(t,r)E_{ab}\partial^{b}Y_{\ell m}(\theta,\varphi),\\
 & h_{ra}=\sum_{\ell, m}h_{1,\ell m}(t,r)E_{ab}\partial^{b}Y_{\ell m}(\theta,\varphi),\\
 & h_{ab}=\frac{1}{2}\sum_{\ell, m}h_{2,\ell m}(t,r)\left[E_{a}^{~c}\nabla_{c}\nabla_{b}Y_{\ell m}(\theta,\varphi)+E_{b}^{~c}\nabla_{c}\nabla_{a}Y_{\ell m}(\theta,\varphi)\right],
\end{aligned}
\end{cases}
\]
where $a,b = \theta,\varphi$, $E_{ab} = \sqrt{\det \gamma}\: \epsilon_{ab}$, with $\gamma_{ab} = \mbox{diag}(1, \;\sin^2\theta)$;
$\epsilon_{ab}$ is
totally antisymmetric symbol ($\epsilon_{\theta\varphi} = 1$) and $\nabla_a$ is covariant derivative on a 2-sphere. Note that perturbations with
$\ell=0$ do not exist and modes with $\ell=1$ are pure gauges~\cite{Armendariz,Kobayashi:odd}. So below in this section we consider perturbations with $\ell \geq 2$. 
% Moreover, in all the relevant expressions below the summation over $\ell$ and $m$ is implicit.

We make use of the gauge freedom and set $h_{2,\ell}=0$ (Regge-Wheeler gauge), so that the quadratic action in the odd-parity sector is derived in terms of $h_{0,\ell m}$ and $h_{1,\ell m}$. It was explicitly shown in Refs.~\cite{Kobayashi:odd,wormhole1} that $h_{0,\ell m}$ is a non-dynamical DOF and one can integrate it out from the quadratic action. The resulting action for the only dynamical DOF reads 
(see Ref.~\cite{wormhole1}
for a detailed derivation):
% \marginpar{integrated over angles?}
\[
\label{eq:action_odd_final}
\begin{aligned}
S^{(2)}_{odd} = \int \mbox{d}t\:\mbox{d}r\:\sqrt{\frac{A}{B}} J^2
\frac{\ell(\ell+1)}{2(\ell-1)(\ell+2)}\cdot \frac{B}{A}\left[ \frac{\mathcal{H}^2}{A \mathcal{G}} \dot{Q}^2 - \frac{B \mathcal{H}^2}{\mathcal{F}} (Q')^2 -\frac{l(l+1)}{J^2}\cdot \mathcal{H} Q^2 - V(r) Q^2 \right] \; ,
\end{aligned}
\]
where an overdot stands for a time derivative, $A$, $B$ and $J$ are metric functions~\eqref{eq:backgr_metric}, $Q$ is an auxiliary field and
\footnote{We have introduced different notations for $\mathcal{H}$ and $\mathcal{G}$ in action~\eqref{eq:action_odd_final} despite their equality for the quadratic subclass of beyond Horndeski theory~\eqref{eq:lagrangian} as these coefficients become different as soon as one adds the cubic subclass to the set-up~\cite{Kobayashi:odd,wormhole1}. } 
\begin{equation}
\label{eq:cal_FGH}
{\cal F}=2 G_4 ,
\qquad
{\cal H}={\cal G}= 2 \left[G_4 - 2X G_{4X} + 4 X^2 F_4 \right].
\end{equation} 
Once $Q$ is known both $h_{0,\ell m}$ and $h_{1,\ell m}$ can be restored.
Note that the third term in eq.~\eqref{eq:action_odd_final} is the 
angular part of the Laplace operator, which governs stability in the angular direction. The "potential" $V(r)$ in eq.~\eqref{eq:action_odd_final} reads
\begin{equation}
\label{eq:Vr}
\begin{aligned}
V(r) &=  \frac{B {\cal H}^2 }{{2\cal F}}
\left[\frac{{\cal F}'}{{\cal F}} \left(2\frac{{\cal H}'}{{\cal H}} -\frac{A'}{A}+\frac{B'}{B}+ 4\frac{J'}{J}\right) -  \frac{{\cal H}'}{{\cal H}} \left( -\frac{A'}{A}+3 \frac{B'}{B}+4 \frac{J'}{J}\right)\right.\\
&\left. - \left(\frac{A'^2}{A^2} -\frac{A'}{A} \frac{B'}{B} - \frac{A''}{A} +\frac{B''}{B} +4 \frac{B'}{B} \frac{J'}{J}-4 \frac{J'^2}{J^2}+ 4 \frac{J''}{J} \right)  - \frac{4}{J^2 B}\cdot \frac{\cal F}{\cal H} \right],
\end{aligned}
\end{equation}
and governs the possible "slow" tachyonic instabilities, which we do not discuss in this paper. However, corresponding constraints following from the potential $V(r)$ may not be particularly restrictive~\cite{Trincherini}.

The quadratic action~\eqref{eq:action_odd_final} gives the following set of stability conditions which ensure the absence of both ghost and gradient instabilities in the odd-parity sector:
% \begin{eqnarray}
% \label{eq:stability_G}
% \mbox{No ghosts:} &\mathcal{G} > 0, \\
% \label{eq:stability_F}
% \mbox{No radial gradient instabilities:} &\mathcal{F} > 0, \\
% \label{eq:stability_H}
% \mbox{No angular gradient instabilities:} &\mathcal{H} > 0.
% \end{eqnarray}
\begin{subequations}
\label{eq:stability_odd}
\begin{align}
\label{eq:stability_G}
\mbox{No ghosts:}\quad &\mathcal{G} > 0, \\
\label{eq:stability_F}
\mbox{No radial gradient instabilities:}\quad &\mathcal{F} > 0, \\
\label{eq:stability_H}
\mbox{No angular gradient instabilities:}\quad &\mathcal{H} > 0.
\end{align}
\end{subequations}
% The sound speed squared for perturbations propagating in the radial and angular directions read, respectively:
% \begin{equation}
% \label{eq:speed_odd}
% c_r^2 = \frac{\mathcal{G}}{\mathcal{F}}, \quad c_{\theta}^2 = \frac{\mathcal{G}}{\mathcal{H}}.
% \end{equation}
To have both radial and angular modes propagating at (sub)luminal speed one has to ensure that corresponding sound speeds squared are not greater than the speed of light
\begin{equation}
\label{eq:speed_odd}
c_r^2 = \frac{\mathcal{G}}{\mathcal{F}} \leq 1, \quad c_{\theta}^2 = \frac{\mathcal{G}}{\mathcal{H}} \leq 1,
\end{equation}
which is the case provided that
\begin{equation}
\label{eq:sublum_odd}
\mathcal{F} \ge \mathcal{G} > 0, \qquad 
\mathcal{H} \ge \mathcal{G} > 0 \; .
\end{equation}
% Constraints in eqs.~\eqref{eq:stability_G}--\eqref{eq:sublem_odd} give the set of stability conditions for the odd-parity sector we use when constructing a stable wormholes solution.

%%%%%%%%%%%%%%%%%%%%%%%%%%%%%%%%%%%%%%%%%%%%%%%%%%%%%%%%%%%%%%%%%%%%%%%
\subsection{Even-parity sector}
\label{sec:even_parity}

We now move on to the even-parity sector, which is the main source of stability constraints for a spherically-symmetric solution.
We adopt the following general form of parametrization for
metric perturbations:
\[
\label{eq:even_parity}
\mbox{Parity~even}\quad \begin{cases}
\begin{aligned}
h_{tt}=&A(r)\sum_{\ell, m}H_{0,\ell m}(t,r)Y_{\ell m}(\theta,\varphi), \\
h_{tr}=&\sum_{\ell, m}H_{1,\ell m}(t,r)Y_{\ell m}(\theta,\varphi),\\
h_{rr}=&\frac{1}{B(r)}\sum_{\ell, m}H_{2,\ell m}(t,r)Y_{\ell m}(\theta,\varphi),\\
h_{ta}=&\sum_{\ell, m}\beta_{\ell m}(t,r)\partial_{a}Y_{\ell m}(\theta,\varphi), \\
h_{ra}=&\sum_{\ell, m}\alpha_{\ell m}(t,r)\partial_{a}Y_{\ell m}(\theta,\varphi), \\
h_{ab}=&\sum_{\ell, m} K_{\ell m}(t,r) g_{ab} Y_{\ell m}(\theta,\varphi)+\sum_{\ell, m} G_{\ell m}(t,r) \nabla_a \nabla_b Y_{\ell m}(\theta,\varphi)\,.
\end{aligned}
\end{cases}
\]
The scalar field $\pi$ also acquires non-vanishing perturbations 
in the parity even sector:
\[
\label{chi}
\pi (t,r,\theta,\varphi) = \pi(r) + \sum_{\ell, m}\chi_{\ell m}(t,r)Y_{\ell m}(\theta,\varphi),
\]
where $\pi(r)$ is the spherically symmetric background field. As we have
mentioned above modes with different $\ell$ and $m$ does not get mixed at the linear level, so to simplify the expressions we usually drop both subscripts in what follows.

The next natural step is to either work with gauge-invariant variables~\cite{Gerlach:1979rw} or to fix the gauge. Below we choose the latter option. Under the infinitesimal coordinate change $x^{\mu} \to x^{\mu} + \xi^{\mu}$ with $\xi^{\mu}$ parametrized as
\begin{equation}
\label{eq:xi}
\xi^{\mu} = \Big(T_{\ell m}(t,r), R_{\ell m}(t,r), \Theta_{\ell m}(t,r) \partial_{\theta}, \dfrac{\Theta_{\ell m}(t,r) \partial_{\varphi}}{\sin^2\theta}\Big)\,  Y_{\ell m}(\theta,\varphi )
\end{equation}
the metric perturbations in eq.~\eqref{eq:even_parity} transform as follows
\begin{equation}
\begin{aligned}
\label{eq:gauge_laws}
H_0 &\rightarrow H_0 + \dfrac{2}{A} \dot{T} - \dfrac{A'}{A} B R, \\
H_1 &\rightarrow H_1 + \dot{R} + T' - \dfrac{A'}{A} T, \\
H_2 &\rightarrow H_2 + 2 B R' + B' R, \\
\beta &\rightarrow \beta + T + \dot{\Theta}, \\
\alpha &\rightarrow \alpha + R + \Theta' - 2 \dfrac{J'}{J} \Theta,\\
K &\rightarrow K + 2 B \dfrac{J'}{J} R, \\
G &\rightarrow G + 2 \Theta, \\
\chi &\rightarrow \chi + B \pi' R,
\end{aligned}
\end{equation}
where we have dropped the arguments of functions for clarity. 
Fixing the gauge amounts to choosing specific functions $T$, $R$ and $\Theta$ in eq.~\eqref{eq:gauge_laws} so that 
some of the perturbations vanish. 
% By means of choosing
% specific functions $T$, $R$ and $\Theta$ in eq.~\eqref{eq:gauge_laws} one can make different sets of perturbations to vanish, i.e. make different gauge choices. 
For instance, 
the original Regge--Wheeler gauge, used e.g. in Refs.~\cite{ReggeWheeler,Armendariz}, amounts to setting $\alpha=\beta=G=0$. 
% in eq.~\eqref{eq:even_parity}. 
Another possible gauge choice was adopted 
in Refs.~\cite{Kobayashi:even,wormhole1} where $\beta=K=G=0$ (let us call it a "spherical gauge" for brevity). Alternatively, Ref.~\cite{Trincherini} used the Regge--Wheeler--unitary gauge with $\beta=G=\chi=0$. 

% Let us note that according to eq.~\eqref{eq:even_parity} fixing the gauge for modes with $\ell = 0$ and 
% $\ell = 1$ requires separate analysis, which we put in Appendix. 
% % since for the monopole mode $\alpha=\beta=G$ automatically, while for the dipole mode $K$ and $G$ are present only in a combination $K-G$
% \marginpar{say elsewhere?}

In our previous work in Ref.~\cite{wormhole1} we have already 
derived the 
action for perturbations in beyond Horndeski theory adopting 
the spherical gauge $\beta=K=G=0$ and found that for modes 
with large $\ell$ the coefficients in the action inevitably become singular at some point. This singularity is unavoidable as soon as one requires the background to be free of ghost instabilities at all points. We provide a detailed explanation in Sec.~\ref{sec:old_gauge}.
% \marginpar{clarify the problem with old gauge!}

There is another subtlety with the gauge choice where $K=0$. 
% in the context of analysing linear stability of 
% spherically-symmetric wormholes. 
Namely, according to the 
transformation laws~\eqref{eq:gauge_laws} 
it is impossible to make $K$ vanishing at all points if
% will reappear if
the metric function $J(r)$ is not a monotonous function, i.e. $J'(r)=0$ at some point. And this is exactly what happens in a wormhole throat, 
since function $J(r)$ describes the profile of a wormhole 
and $J'(r)=0$ at $r=R_{min}$~\eqref{eq:metric_bounded}. Hence, 
the spherical gauge in Ref.~\cite{wormhole1} is 
not entirely convenient for checking linear stability 
of the wormhole solutions near the throat.

In this paper we aim to formulate a complete set of stability conditions 
for modes with high momentum, which does not involve any singularities and, hence, is suitable for stability analysis throughout the whole space, where the wormhole is present. 

% In this paper we aim to not only construct a wormhole solution and check its stability in a more appropriate gauge, but also
% find the gauge which has no singularity problem in the quadratic action mentioned above.   

% In the following two subsections we aim to derive the quadratic action for even-parity perturbations in two different gauges: 1) $\beta=K=G=0$ (adopted in Refs.~\cite{wormhole1}) and 2) Regge--Wheeler--unitary gauge with $\beta=G=\chi=0$. In our previous work in Ref.~\cite{wormhole1} we have already calculated the 
% action for perturbations in the gauge $\beta=K=G=0$ (let us call it the "old" gauge from now on for brevity) and found that for high momenta modes with large $\ell$ the action is inevitably singular at some point. This singularity is inevitable as soon as one requires the background to be free of ghost instabilities. 

In Sec.~\ref{sec:old_gauge} we give a schematic derivation of the action for even-parity perturbations 
for beyond Horndeski Lagrangian~\eqref{eq:lagrangian} 
in the spherical gauge (see Ref.~\cite{wormhole1} for a detailed procedure)
and discuss its drawbacks in the context of stability analysis for spherically-symmetric solutions like wormholes. In Sec.~\ref{sec:new_gauge} we
calculate the quadratic action in Regge--Wheeler--unitary gauge and  find out if we can avoid singularities in the coefficients of quadratic action analogous to those in the old gauge. In result, we derive the whole set of stability conditions for modes with high $\ell$.

%%%%%%%%%%%%%%%%%%%%%%%%%%%%%%%%%%%%%%%%%%%%%%%%%%%%%%%%%%%%%%%%%%%%%%%%%
\subsubsection{Existing results in the spherical gauge ($\beta=K=G=0$)}
\label{sec:old_gauge}
The quadratic action for even-parity perturbations~\eqref{eq:even_parity}
with $\ell \geq 2$ 
\footnote{Both $\ell = 0$ and $\ell=1$ are somewhat special cases, which, however, do not give anything particularly new for stability analysis as compared to $\ell \geq 2$~\cite{Kobayashi:even,wormhole1}.}
and gauge conditions $\beta=K=G=0$
reads:
\begin{equation}
\label{eq:action_even}
\begin{aligned}
&S_{even}^{(2)} = \int \mbox{d}t\:\mbox{d}r \left(H_0 \left[ a_1 \chi''+ a_2
\chi'+a_3 H_2'+j^2 a_4 \alpha'+\left( a_5+j^2 a_6 \right) \chi\right.\right.  \\
&\left.+\left( a_7+j^2 a_8 \right)H_2+j^2a_9 \alpha \right]
%%%%%
+j^2 b_1 H_1^2+H_1 \left[b_2 {\dot {\chi}}'+b_3 {\dot {\chi}}+b_4 {\dot H_2}+j^2b_5 {\dot \alpha}\right] \\
&+ c_1{\dot H_2} {\dot {\chi}} + H_2 \left[c_2  \chi'+\left( c_3+j^2 c_4\right)\chi + j^2 c_5 \alpha \right] + c_6 H_2^2+j^2d_1 {\dot \alpha}^2
\\
&\left.+j^2 d_2 \alpha\chi'+j^2 d_3 \alpha\chi+j^2 d_4 \alpha^2
+ e_1{\dot {\chi}}^2+e_2 \chi'^2+\left( e_3+j^2 e_4  \right) \chi^2\right),
\end{aligned}
\end{equation}
where the subscripts $\ell$, $m$ are again omitted, $j^2=\ell(\ell+1)$, and
we have integrated over $\theta$ and $\phi$. The
explicit expressions for coefficients $a_i$, $b_i$, $c_i$, $d_i$
and $e_i$ with $\sqrt{-g}$ included are given in Appendix B.

It follows from action~\eqref{eq:action_even} that $H_0$ is a Lagrange multiplier and $H_1$ is also a non-dynamical DOF, which give the following constraints, respectively:
\begin{equation}
\label{eq:H0old}
\begin{aligned}
a_1 \chi''+a_2  \chi'+a_3 H_2'+
j^2 a_4 \alpha'+\left( a_5+j^2 a_6 \right) \chi
+\left( a_7+j^2 a_8 \right)H_2+j^2a_9 \alpha = 0,
\end{aligned}
\end{equation}
% and $H_1$ is also a non-dynamical DOF, giving the constraint
\begin{equation}
\label{eq:H1old}
 H_1 = - \frac{1}{2 j^2 b_1} \left( b_2 {\dot {\chi}}'+b_3 {\dot {\chi}}+b_4 {\dot H_2}+j^2b_5 {\dot \alpha}\right).
\end{equation}
Introducing a new variable $\psi$ as
\begin{equation}
\label{eq:H2old}
H_2 = \psi - \frac{1}{a_3}\left( a_1 \chi' + j^2 a_4 \alpha\right),
\end{equation}
and so that upon substitution of $H_2$ into eq.~\eqref{eq:H0old}
both $\chi''$ and $\alpha'$ get cancelled out automatically, the resulting equation becomes algebraic for $\alpha$ and one can express it in terms of $\psi$ and $\chi$ 
\begin{equation}
\label{eq:alphaold}
\alpha = \frac{a_3^2 \psi' + a_3(a_7+j^2a_8) \psi + [a_3 (a_2 - a_1') -j^2 a_1 a_8] \chi' + a_3 (a_5 + j^2 a_6) \chi}{j^2 \left[ a_3 a_4' - a_3' a_4 - a_3 a_9 + a_4 (a_7 + j^2 a_8\right]} \;.
\end{equation}
Then it becomes possible to express both $H_2$ in eq.~\eqref{eq:H2old} and $H_1$ in eq.~\eqref{eq:H1old} in terms 
of $\psi$ and $\chi$ only. Hence, the quadratic action~\eqref{eq:action_even} can be cast in terms of two dynamical DOFs:
\[
\label{eq:even_action_final_old}
S_{even}^{(2)} = \int \mbox{d}t\:\mbox{d}r \sqrt{\frac{A}{B}} J^2 \left(
\frac12 \mathcal{K}_{ij} \dot{v}^i \dot{v}^j - \frac12 \mathcal{G}_{ij} v^{i\prime} v^{j\prime} - \mathcal{Q}_{ij} v^i v^{j\prime} - \frac12 \mathcal{M}_{ij} v^i v^j \right),
\]
where $i=1,2$ and $v^1=\psi$, $v^2=\chi$. We note that
terms which are higher order in derivatives, like
$\dot{\psi}' \dot{\chi}$, disappear upon integrating by parts.

The no-ghost condition amounts to requiring that $\mathcal{K}_{ij}$ in eq.~\eqref{eq:even_action_final_old} is positive definite:
\[
\label{eq:no_ghost_old}
\mathcal{K}_{11}>0, \qquad \det(\mathcal{K}) > 0,
\]
where
\begin{equation}
\label{eq:K11old}
{\cal K}_{11}=\frac{8  B {\left( 2{\cal H} J J' +
\Xi \pi' \right)}^2 \left[ \ell (\ell+1){\cal P}_1-{\cal F}\right]}
{\ell (\ell+1) A^2 \mathcal{H}^2  \TT^2},
\end{equation}
\begin{equation}
\label{eq:detKold}
\det ({\cal K})=\frac{16 B J'^2 (\ell-1) (\ell+2){\left( 2{\cal H} J J' + \Xi \pi'\right)}^2 \left[{\cal F}(2{\cal P}_1-{\cal F})\right]}
{\ell (\ell+1) A^3 J^2 \pi'^2 \mathcal{H}^2 \TT^2},
\end{equation}
with $\mathcal{F}$ and $\mathcal{H}$ are given by \eqref{eq:cal_FGH},
\[
  \begin{aligned}
\label{eq:Xi}
\Xi =  2G_{4\pi}J^2 + 4G_{4X}BJJ'\pi'
 -2G_{4\pi X}BJ^2\pi'^2
  % \\&
 - 4G_{4XX}B^2JJ'\pi'^3 
   + 16F_{4}B^2JJ'\pi'^3 - 4F_{4X}B^3JJ'\pi'^5, \\
    \end{aligned}
\]
and
\begin{subequations}
\begin{align}
\label{eq:Theta}
\TT &= 2 \ell (\ell+1) \: J\left(\mathcal{H} - 2F_4 B^2 \pi'^4\right) +\mathcal{P}_2,\\
\label{eq:P1}
\mathcal{P}_1 &= \frac{\sqrt{B}}{\sqrt{A}} \cdot
\frac{\mbox{d}}{\mbox{d}r}\left[\frac{\sqrt{A}}{\sqrt{B}}
\frac{J^2 \mathcal{H}\left(\mathcal{H} - 2 F_4 B^2 \pi'^4\right)}{2{\cal H} J J' + \Xi \pi'}\right],\\
\mathcal{P}_2 &= \frac{B(A' J - 2 A J')}{A}
\left(2{\cal H} J J' + \Xi \pi'\right)
\; .
\end{align}
\end{subequations}
Here and in what follows we highlight $\TT$ in bold to emphasize that it involves $\ell(\ell+1)$. It follows immediately form eqs.~\eqref{eq:K11old}-\eqref{eq:detKold}
that both no-ghost conditions~\eqref{eq:no_ghost_old} are satisfied 
provided
\[
\label{eq:stability_even_ghost}
2{\cal P}_1-{\cal F} > 0.
\]

To have no radial gradient instabilities one also has to require positive definiteness for matrix $\mathcal{G}_{ij}$ in eq.~\eqref{eq:even_action_final_old}:
\[
\label{eq:no_gradient_old}
\mathcal{G}_{11}>0,  \qquad \det{\mathcal{G}}>0.
\]
Here
\begin{equation}
\label{eq:G11old}
\mathcal{G}_{11}=\frac{4 B^{2} \left[ \mathcal{G} (\ell+2)(\ell-1)  (2 \mathcal{H} J J'+ \Xi \pi')^2+\ell(\ell+1)(2 J^2 \Gamma \mathcal{H} \Xi \pi'^2  - \mathcal{G} \Xi^2 \pi'^2-4 J^4 \Sigma \mathcal{H}^2/B )\right]}
{ \ell (\ell+1) A \mathcal{H}^2  \TT ^2},
\end{equation}
\begin{equation}
\label{eq:detGold}
\det(\mathcal{G})=\frac{16  B^3 J'^2 \mathcal{G} (\ell-1) (\ell+2) (2 J^2 \Gamma \mathcal{H} \Xi \pi'^2  - \mathcal{G} \Xi^2 \pi'^2-4 J^4 \Sigma \mathcal{H}^2/B )}
{ \ell (\ell+1) A  J^2 \pi'^2  \mathcal{H}^2 \TT^2},
\end{equation}
with
\begin{align}
\label{eq:Gamma}
& \Gamma = \Gamma_1 + \frac{A'}{A} \Gamma_2,
\\
& \nonumber\Gamma_{1} = 4\left(G_{4\pi} + 2XG_{4\pi X} + \frac{B\pi'J'}{J}(G_{4X} + 2XG_{4XX}) \right) - \frac{16J'}{J}B\pi'X(2F_{4} + XF_{4X}), \nonumber\\
&\Gamma_{2} = 2B\pi'\left(G_{4X} - B\pi'^2G_{4XX}\right)
- 8B\pi'X(2F_{4} + XF_{4X}),
\nonumber
\end{align}
and
\begin{align}
\label{eq:KSI}
& \Sigma = XF_{X} + 2F_{XX}X^2 +2\left(\frac{1 - BJ'^2}{J^2} - \frac{BJ'}{J} \frac{A'}{A}\right)X(G_{4X} + 2XG_{4XX})
\nonumber\\& 
     - \frac{4BJ'}{J}\left( \frac{J'}{J } + \frac{A'}{A}\right)X^2(3G_{4XX} + 2XG_{4XXX}) +  2B\pi'\left(\frac{4J'}{J } + \frac{ A'}{A} \right)X( \frac{3}{2} G_{4\pi X} + XG_{4 \pi XX})
      \nonumber\\& 
      + \frac{B^3J'(A'J + AJ')}{AJ^2}\pi'^4(12F_{4} - 9F_{4X}B\pi'^2 + F_{4XX}B^2\pi'^4).
\end{align}
According to eqs.~\eqref{eq:G11old} and~\eqref{eq:detGold} both conditions~\eqref{eq:no_gradient_old} hold provided that
\[
% \label{eq:P3}
\label{eq:stability_even_radial}
2 J^2 \Gamma \mathcal{H} \Xi \pi'^2  - \mathcal{G} \Xi^2 \pi'^2-4 J^4 \Sigma \mathcal{H}^2/B  > 0.
\]

The corresponding sound speeds squared for perturbations propagating along the radial direction are given by the eigenvalues of matrix $(AB)^{-1}(\mathcal{K})^{-1}\mathcal{G}$:
\begin{equation}
\label{eq:speed_even_old}
c_{s1}^2 = \frac{\mathcal{G}}{\mathcal{F}},
\qquad
c_{s2}^2 = \frac{(2 J^2 \Gamma \mathcal{H} \Xi \pi'^2  - \mathcal{G} \Xi^2 \pi'^2-4 J^4 \Sigma \mathcal{H}^2/B )}
{\left(2{\cal H} J J' + \Xi \pi'\right)^2 (2{\cal P}_1-{\cal F})} \; ,
\end{equation}
where $c_{s1}^2$ coincides with the radial speed squared $c^2_r$ for odd-parity modes in eq.~\eqref{eq:speed_odd}, which enables us
to interpret it as a propagation speed in the radial direction of two tensor modes. 
% To sum up, satisfying conditions~\eqref{eq:no_ghost_old} and~\eqref{eq:no_gradient_old} ensures there are no ghosts and 
% radial gradient instabilities.  
% To ensure that the even-parity modes propagate at speeds not exceeding that of light

Positive definiteness of matrices $\mathcal{M}_{ij}$ and 
$\mathcal{Q}_{ij}$ in the action~\eqref{eq:even_action_final_old} also provide stability conditions, which ensure that angular gradient instabilities and "slow"
tachyonic modes are absent. This set of conditions was not thoroughly addressed in previous works in Refs.~\cite{Kobayashi:even,wormhole1,Trincherini} and we also omit discussing them for now. 
We explicitly give the matrix $\mathcal{M}_{ij}$ for high angular momentum modes
as well as corresponding stability conditions for angular gradient instabilities in Regge-Wheeler-unitary 
gauge in Sec.~\ref{sec:new_gauge}.

Let us now show that the coefficients in the action~\eqref{eq:even_action_final_old} become singular for modes with 
high momentum $\ell$ if the no-ghost condition~\eqref{eq:stability_even_ghost} is satisfied at all points.
% address the problem of singular coefficients in the action~\eqref{eq:even_action_final_old} mentioned above.
%%
% According to the no-ghost condition~\eqref{eq:no_ghost_old} 
% ${\cal P}_1$ in eq.~\eqref{eq:P1} has to be positive, since
% $\cF > 0$ due to stability constraints in the odd-parity sector~\eqref{eq:sublum_odd}:
%%
% As $\cF > 0$ due to stability constraints in the odd-parity sector~\eqref{eq:sublum_odd}, ${\cal P}_1$ in eq.~\eqref{eq:P1}
% has to be positive as well:
The no-ghost condition~\eqref{eq:stability_even_ghost} together with eq.~\eqref{eq:P1} can be rewritten as follows:
\[
\label{eq:no_go1}
\frac{\sqrt{B}}{\sqrt{A}} \xi' >\frac{\cF}{2} \quad \mbox{with}
\quad \xi = \frac{\sqrt{A}}{\sqrt{B}}
\frac{J^2 \mathcal{H}\left(\mathcal{H} - 2 F_4 B^2 \pi'^4\right)}{2{\cal H} J J' + \Xi \pi'}.
\]
As $\cF > 0$ due to stability constraints in the odd-parity sector~\eqref{eq:stability_odd} it follows from eq.~\eqref{eq:no_go1} that $\xi'$ has to be positive.
% It follows from eq.~\eqref{eq:no_go1} 
Then $\xi$ is a monotonously growing function, so $\xi=0$ at some point(s)
\footnote{All other options where $\xi'>0$ but $\xi$ does not cross zero involve either fine-tuning or special cases like $\cF \to 0$ as $r \to -\infty$, which signals potential strong coupling in the odd-parity sector, see eq.~\eqref{eq:action_odd_final} (see e.g. Ref.~\cite{wormhole1} for a discussion).}.
The latter happens provided the numerator of $\xi$ crosses zero,
which is possible since the 
combination $\left(\mathcal{H} - 2 F_4 B^2 \pi'^4\right)$ can take zero value, while $\cH >0$ due to stability in the odd-parity sector~\eqref{eq:stability_odd}. 
Let us note here that 
in Horndeski theories $F_4 = 0$, so $\xi$ cannot cross zero in a healthy way and, hence, the no-ghost condition~\eqref{eq:stability_even_ghost} cannot be met at all points in this case. This is what is usually referred to as a no-go theorem in Horndeski theories, see Refs.~\cite{Rubakov:2016zah,Olegi} for details.

What is important for us here is that the same combination 
$\left(\mathcal{H} - 2 F_4 B^2 \pi'^4\right)$ gives the leading contribution into $\TT$ in eq.~\eqref{eq:Theta} for modes with $\ell \gg 1$. 
% i.e. $\TT$ can effectively cross zero for high momenta modes. 
% This in turn means that as soon as one satisfies 
% the no-ghost condition~\eqref{eq:stability_even_ghost}
% by having $\left(\mathcal{H} - 2 F_4 B^2 \pi'^4\right)=0$ 
% at some point(s), one risks 
% to have 
% the components of both matrices $\mathcal{K}_{ij}$ and $\mathcal{G}_{ij}$
% % $\mathcal{K}_{11}$~\eqref{eq:K11old} 
% % and $\det(\mathcal{K})$~\eqref{eq:detKold} 
% hit infinity for high momenta modes since they involve $\TT^2$ in denominator, see eqs.~\eqref{eq:K11old},~\eqref{eq:detKold},~\eqref{eq:G11old} and ~\eqref{eq:detGold}. 
This in turn means that as soon as
the no-ghost condition~\eqref{eq:stability_even_ghost} is satisfied
by having $\left(\mathcal{H} - 2 F_4 B^2 \pi'^4\right)=0$ 
at some point(s), $\TT \to 0$ for high momenta modes. Thus,
the components of both matrices 
$\mathcal{K}_{ij}$ and $\mathcal{G}_{ij}$
hit infinity for high momenta modes since they involve $\TT^2$ in denominators, see eqs.~\eqref{eq:K11old},~\eqref{eq:detKold},~\eqref{eq:G11old} and ~\eqref{eq:detGold}. 
This is the problem
of singular coefficients in the quadratic action upon $\TT$-crossing. 

The appearance of $\TT$ in denominator seems to result from the gauge choice, and this choice, in particular, dictates the way non-dynamical DOFs are integrated out in action~\eqref{eq:action_even}. Namely, the $\ell^2$ part of $\TT$ comes from 
the combination (see Appendix B for definitions of $a_i$)
$$j^2 a_4 a_8 = j^2 \frac A2 \cdot \mathcal{H} \left(\mathcal{H} - 2 F_4 B^2 \pi'^4\right) \;,$$
which arises in the denominator of eq.~\eqref{eq:alphaold} and originates from the field redefinition~\eqref{eq:H2old}. 

The fact that $\TT$-crossing
is directly related to ensuring stability at the linearized level makes the old gauge unsuitable for construction of a completely stable wormhole solution.

In the following section we rederive the quadratic action for
beyond Horndeski Lagrangian~\eqref{eq:lagrangian} 
in a Regge-Wheeler-unitary gauge and pay special attention to potential divergencies in the coefficients of the action due to $\TT$ crossing zero. So we keep track of $\TT$ and ensure the coefficients in the quadratic action do not hit singularities, so that we end up with  
a regular quadratic action.

%%%%%%%%%%%%%%%%%%%%%%%%%%%%%%%%%%%%%%%%%%%%%%%%%%%%%%%%%%%%%%%%%%%%%%%%
\subsubsection{Original calculations in the Regge-Wheeler-unitary gauge ($\beta=G=\chi=0$)}
\label{sec:new_gauge}

The quadratic action for even-parity modes in the Regge-Wheeler-unitary gauge reads:
\begin{equation}
\label{eq:action_even_RW}
\begin{aligned}
&S_{even}^{(2)} = \int \mbox{d}t\:\mbox{d}r 
\left(H_0 \left[ a_3 H_2' + \left( a_7 + j^2 a_8 \right)H_2  + j^2 a_4 \alpha' + j^2 a_9 \alpha + a_{13} K'' + a_{14} K'\right.\right.  \\
&\left.\left.  
+ (a_{15} + j^2 a_{16}) K \right]
%%%%%
+j^2 b_1 H_1^2 + H_1 \left[b_4 {\dot H_2}+j^2b_5 {\dot \alpha} + b_{8}\dot{K}' + b_{9}\dot{K} \right] + c_{8} \dot{H}_2 \dot{K} + c_6 H_2^2 \right. \\
&\left. + H_2 \left[ j^2 c_5 \alpha + c_{11} K' + (c_{12} +j^2 c_{13})K \right]   + j^2d_1 {\dot \alpha}^2 +j^2 d_4 \alpha^2 +  j^2 d_{7} \alpha K'  + p_{8}\dot{K}^2 + p_{9}K'^2
\right),
\end{aligned}
\end{equation}
where the notations are similar to those in action~\eqref{eq:action_even}
and the coefficients $a_i$, $b_i$, $c_i$, $d_i$ and $p_i$ are given in Appendix B. In full analogy with the spherical gauge case above variation of the action~\eqref{eq:action_even_RW} w.r.t. $H_0$ and $H_1$ gives two constraint equations, respectively:
\begin{equation}
\label{eq:H0_RW}
a_3 H_2' +\left( a_7+j^2 a_8 \right)H_2 + j^2 a_4 \alpha'
+ j^2 a_9 \alpha + a_{13}K'' + a_{14}K' +(a_{15}+j^2 a_{16})K = 0,
\end{equation}
\begin{equation}
\label{eq:H1_RW}
H_1 = - \frac{1}{2 j^2 b_1} \left( b_4 {\dot H_2}+j^2b_5 {\dot \alpha} 
  + b_8 \dot{K}' +b_9 \dot{K} \right).
\end{equation}
This time the constraint~\eqref{eq:H0_RW} contains not only the first derivatives of $H_2$, $\alpha$ and $K$ but also $K''$. To simplify things
let us introduce an axillary field $\psi$ as follows (cf. eq.~\eqref{eq:H2old})
\begin{equation}
\label{eq:alpha_shifted}
\alpha = \psi - \dfrac{a_{13}}{j^2 a_4}  K'. 
\end{equation}
Let us note here that $a_{13}= J^2 \cdot a_4$ (see Appendix B), so the redefinition above is regular.
Upon substitution of $\alpha$ and $\alpha'$ from eq.~\eqref{eq:alpha_shifted} into the constraint~\eqref{eq:H0_RW}, one can see that not only $K''$ get canceled out but also the resulting factor in front of $K'$ is automatically vanishing (see Appendix B for the explicit form of $a_i$):
\[
a_{14} - a_{13}' +\frac{a_{13}}{a_4}(a_4' - a_9) = 0.
\]
Hence, the resulting equation is algebraic for $K$ and reads
\begin{equation}
\label{eq:K}
K = - \dfrac{1}{a_{15} +j^2 a_{16}} \left[ a_3 H_2' +(a_7+j^2 a_8) H_2 +j^2 a_4 \psi' +j^2 a_9 \psi \right].
\end{equation}
Now by substituting $\alpha$ from eq.~\eqref{eq:alpha_shifted} and $K$ from eq.~\eqref{eq:K} into the second constraint~\eqref{eq:H1_RW} one can express $H_1$ in terms of $H_2$ and $\psi$. Therefore, by making use of both constraint equations~\eqref{eq:H0_RW}-\eqref{eq:H1_RW} and field redefinition~\eqref{eq:alpha_shifted} one can cast the quadratic action~\eqref{eq:action_even_RW} in terms of $H_2$ and $\psi$:
\begin{equation}
\label{eq:unconstrained_action_RW}
\begin{aligned}
S_{even}^{(2)} = &\int \mbox{d}t\:\mbox{d}r \left(
g_{4(i)}\dot{\psi}'\:^2 + g_{5(i)}\psi''\:^2 
+f_{4(i)}\dot{H_2}'^2  + f_{5(i)} H_2''^2
+ h_{7(i)} \dot{H_2}' \dot{\psi}' +  h_{8(i)} H_2'' \psi''
\right.\\ 
&\left.
+g_{1(i)}\dot{\psi}^2 + g_{2(i)}\psi'\:^2 +  g_{3(i)}\psi^2
 + f_{1(i)}\dot{H_2}^2 + f_{2(i)}H_2'^2 + f_{3(i)}H_2^2 
+ h_{5(i)}\dot{H_2} \dot{\psi}' 
\right.\\ 
&\left.
+ h_{6(i)} H_2' \psi'' + h_{1(i)}\dot{H_2} \dot{\psi}  + h_{2(i)} H_2' \psi' + h_{3(i)} H_2 \psi'  + h_{4(i)} H_2 \psi
\right),
\end{aligned}
\end{equation}
where $g_{j(i)}$, $f_{j(i)}$ and $h_{j(i)}$ are some combinations of the initial coefficients $a_i$, $b_i$, $c_i$, $d_i$ and $p_i$ in eq.~\eqref{eq:action_even_RW} and we give explicitly only those attributed to terms with four derivatives (see the first line of eq.~\eqref{eq:unconstrained_action_RW}): 
\begin{equation}
\label{eq:action_coef1}
\begin{aligned}
& g_{4(i)} = \frac{2\ell(\ell+1)B^{3/2} J^2 }{(\ell+2)(\ell-1)A^{1/2}}\frac{\cH^2}{\cF}, \qquad \qquad
g_{5(i)}= - g_{4(i)} \cdot A B \frac{\cG}{\cF}, \\
& f_{4(i)} = \frac{B^{3/2} J^2}{2 \ell(\ell+1)(\ell+2)(\ell-1) A^{1/2}} \frac{(2\cH J J' + \Xi \pi')^2}{\cF}, \quad
f_{5(i)} = - f_{4(i)} \cdot A B \frac{\cG}{\cF}, \\
& h_7 = - \frac{2 B^{3/2} J^2}{(\ell+2)(\ell-1) A^{1/2}} \frac{\cH(2\cH J J' + \Xi \pi')}{\cF}, \qquad \qquad
h_8 = - h_7 \cdot A B \frac{\cG}{\cF},
\end{aligned}
\end{equation}
where we made use of Appendix B to substitute $a_i$, $b_i$, etc. The rest of coefficients in eq.~\eqref{eq:unconstrained_action_RW} are irrelevant at this point, however, we should note that all of them are regular.
Note that according to
the unconstrained action~\eqref{eq:unconstrained_action_RW} 
both $\psi$ and $H_2$ are naively described by the fourth order equations. Let us now show that in fact both DOFs are described by the second order equations upon making suitable field redefinitions.

It immediately follows from eq.~\eqref{eq:action_coef1} that the following relations hold
\begin{equation}
\begin{aligned}
&4\; g_{4(i)} \cdot f_{4(i)} - h_7^2 = 0, \\
&4\;g_{5(i)} \cdot f_{5(i)} - h_8^2 = 0,
\end{aligned}
\end{equation}
% \marginpar{\bf check!}
which means that the corresponding terms $\dot{\psi}'^2$, $\dot{H_2}'^2$, $\dot{H_2}'\dot{\psi}'$ and $\psi''^2$, $H_2''^2$, $\psi''H_2''$ get combined into perfect squares in eq.~\eqref{eq:unconstrained_action_RW}. Hence, upon a field redefinition
\begin{equation}
\label{eq:tildePsi}
\begin{aligned}
& \psi = \tpsi + \frac{(2\cH J J' + \Xi \pi')}{2\ell(\ell+1) \; \cH} \tH, \\
& H_2 = \tH.
\end{aligned}
\end{equation}
the terms $\dot{\tilde{H_2}}'^2$, $\tilde{H_2}''^2$, $\dot{\tilde{H_2}}' \dot{\psi}'$ and $\tilde{H_2}'' \psi''$ vanish and the action~\eqref{eq:unconstrained_action_RW} reduces to the following:
\begin{equation}
\label{eq:unconstrained_action_tildePsi}
\begin{aligned}
S_{even}^{(2)} = &\int \mbox{d}t\:\mbox{d}r \left(
g_4\; \dot{\tilde\psi}'\:^2  + g_5 \;\tilde\psi''\:^2 + g_1\; \dot{\tilde\psi}^2 + g_{2} \;\tilde\psi'\:^2 + g_{3} \;\tilde\psi^2
+ f_1 \;\dot{\tilde{H_2}}^2 + f_2 \;\tilde{H_2}'^2 \right.\\ 
&\left.+ f_{3} \;\tilde{H_2}^2 
+ h_5\; \dot{\tilde{H_2}} \dot{\tilde\psi}' + h_6\; \tilde{H_2}' \tilde\psi'' + h_1\; \dot{\tilde{H_2}} \dot{\tilde\psi}  + h_2\; \tilde{H_2}' \tilde\psi'  + h_{3}\; \tilde{H_2} \tilde\psi'+ h_{4}\; \tilde{H_2} \tilde\psi
\right),
\end{aligned}
\end{equation}
where 
\begin{equation}
\label{eq:coef_tildePsi}
\begin{aligned}
& g_4 = \frac{2 \ell(\ell+1) B^{3/2} J^2 }{(\ell^2+\ell-2) A^{1/2}} \frac{\cH^2}{\cF}, 
\qquad \qquad
g_5 = - \frac{2 \ell(\ell+1) A^{1/2} B^{5/2} J^2}{\ell^2 + \ell -2} \frac{\cH^3}{\cF^2}, \\
& 
h_5 = - \frac{B^{1/2} J}{(\ell^2+\ell-2) A^{1/2}} \frac{\cH \TT}{\cF},
\qquad \qquad
h_6 = \frac{A^{1/2}B^{3/2} J}{(\ell^2+\ell-2)} \frac{\cH^2 \TT}{\cF^2},\\
&f_1 = \frac{1}{8\ell(\ell+1)(\ell^2+\ell-2)(AB)^{1/2}} \frac{\TT^2}{ \cF},
\qquad \qquad
f_2 = \frac{  (AB)^{1/2}}{8\ell(\ell+1)(\ell^2+\ell-2)}\frac{\cH \TT^2}{ \cF^2},
\end{aligned}
\end{equation}
while the explicit form of the rest of coefficients 
$g_j$, $f_j$ and $h_j$ is irrelevant here (they do not involve $\TT$ and were checked to be perfectly regular).
% are given in Appendix C and all the coefficients 
% are still perfectly regular.
Note that there are still terms with higher order derivatives 
of $\tpsi$ in the
action~\eqref{eq:unconstrained_action_tildePsi}. 
However, we see again
that terms $g_4\; \dot{\tilde\psi}'\:^2 $, $h_5\; \dot{\tilde{H_2}}\dot{\tilde\psi}'$, $f_1 \dot{\tilde{H_2}}^2$ and $g_5 {\tilde\psi}''\:^2$,
$h_6\; \tilde{H_2}'{\tilde\psi}''$, $f_2\;\tilde{H_2}'^2$ get combined into perfect squares since
% \marginpar{\bf check!}
\begin{equation}
\label{eq:pefect_squares2}
\begin{aligned}
&4\; g_{4} \cdot f_{1} - h_5^2 = 0, \\
&4\;g_{5} \cdot f_{2} - h_6^2 = 0,
\end{aligned}
\end{equation}
hence, it is possible to redefine $\tH$ as
% so that $\dot{\tilde\psi}'^2$, 
% ${\tilde\psi}''^2$, $\tilde{H_2}' \tilde\psi''$ and $\dot{\tilde{H_2}} \dot{\tilde\psi}'$
% get cancelled out in the action~\eqref{eq:unconstrained_action_tildePsi}
% and $\tpsi$ becomes a conventional field decribed by the second order equations:
\begin{equation}
\label{eq:tildePsi2}
\begin{aligned}
& \tpsi = \bpsi, \\
& \tH = \bH + 4 \ell(\ell+1) B J \dfrac{\;{\cal H} }{\TT} \; \bar\psi'.
\end{aligned}
\end{equation}
Then the terms with higher derivatives vanish and the resulting action gives the second order differential equations of motion for both $\bpsi$ and $\bar{H_2}$ in full analogy with the case of Sec.~\ref{sec:old_gauge}. 

There is, however, a problem with redefinition~\eqref{eq:tildePsi2}: it diverges in the vicinity of $\TT = 0$, which means that the final action for $\bpsi$ and $\bar{H_2}$
also involves singular coefficients and this is exactly what we tried to avoid. One possible way to proceed is to refrain from the redefinition~\eqref{eq:tildePsi2} and use the action~\eqref{eq:unconstrained_action_tildePsi} for further stability analysis. The advantage of the latter approach is that all coefficients in the action are regular everywhere
including $\TT=0$, but at the same time there are still higher derivative terms 
of $\tpsi$, which naively signal that the corresponding system of equations of motion
might admit more than four solutions for $\tpsi$ and $\tilde{H_2}$, i.e. there are additional pathological DOFs. 
One should note, however, that the redefinition~\eqref{eq:tildePsi2} is perfectly fine away from $\TT=0$ and it demonstrates that in fact the action~\eqref{eq:unconstrained_action_tildePsi} gives two second order differential equations, which normally admit four solutions. So in the following section let us
prove that even in the vicinity of $\TT=0$ the action~\eqref{eq:unconstrained_action_tildePsi} describes a healthy system, giving four regular solutions ${\left(
  \begin{array}{c}
    \tpsi \\
    \tilde{H}_2 \\
  \end{array}
\right)}$.
This legitimizes redefinition~\eqref{eq:tildePsi2} at all points including $\TT=0$ since we explicitly show that potential divergencies in the final quadratic action for $\bpsi$
and $\bar{H_2}$ are not inherited from the original action~\eqref{eq:unconstrained_action_tildePsi} and
do not have any physics behind them but result from our field redefinition.

\subsubsection{$\TT$-crossing}
\label{sec:TT_crossing}

% Let us check that both $\tilde{H_2}$ and $\tilde{\psi}$ are regular in the vicinity of $\TT = 0$. 
Let us explicitly solve the linearized equations for $\tilde{H_2}$ and $\tilde{\psi}$ following from the action~\eqref{eq:unconstrained_action_tildePsi} in the vicinity of $\TT=0$.
The general form of these linearized equations reads:
% The action~\eqref{eq:unconstrained_action_tildePsi} gives the following linearized equations in terms of $\tilde{H_2}$ and $\tilde{\psi}$: 
\begin{subequations}
\label{eq:deltaPsiH2}
\begin{align}
\label{eq:deltaPsi}
\delta\tilde{\psi}: \;\; &
 2 g_5 \tpsi'''' + 4 g_5' \tpsi''' + (2 g_5 '' - 2 g_2)\tpsi'' - 2 g_2' \tpsi'  + 2 g_3 \tpsi  + h_6 \tilde{H_2}''' + (2 h_6' - h_2) \tilde{H_2}'' \nonumber\\
& + (h_6'' -  h_2' - h_3) \tilde{H_2}' + (h_4 - h_3')\tilde{H_2} + 2 g_4 \ddot{\tpsi}'' + 2 g_4' \ddot{\tpsi}' 
- 2 g_1 \ddot{\tpsi} + h_5 \ddot{\tH}' + (h_5' - h_1) \ddot{\tH} = 0, \\\nonumber
\\
% \end{equation}
% %%%%%%
% \begin{equation}
\label{eq:deltaH2}
\delta\tilde{H_2}: \;\; &
2 f_2 \tilde{H_2}'' + 2 f_2 \tilde{H_2}' - 2 f_3 \tilde{H_2} + h_6 \tpsi''' + (h_6'+h_2)\tpsi''
+ (h_2' - h_3)\tpsi' - h_4 \tpsi 
+ 2 f_1 \ddot{\tH} + h_5 \ddot{\tpsi}' 
+h_1 \ddot{\tpsi} = 0.
\end{align}
\end{subequations}
% where dots stand for the terms with time-derivatives, e.g. 
% $\ddot{\tilde{\psi}}''$, $\ddot{\tpsi}'$, $\ddot{\tpsi}$, 
% $\ddot{\tilde{H}}_2'$
% and $\ddot{\tilde{H}}_2$, which we have omited here for brevity
% \footnote{In the general case of a time-dependent background one could take these terms into account by making use of a Fourier transformation for the time coordinate. However, these terms do not affect the regular behaviour of solutions as they contribute to the subleading derivative terms $\tpsi''$, $\tilde{H_2}'$, etc.}. 
Since we consider a static, spherically-symmetric background 
for both $\tpsi(t,r)$ and 
$\tilde{H_2}(t,r)$ we use constant energy Ansatz:
\begin{equation}
\tpsi(t,r) = \tpsi(r) e^{i\omega t}, \qquad
\tilde{H_2}(t,r) = \tilde{H_2}(r) e^{i\omega t},
\end{equation}
where $\omega$ denotes frequency. Until the end of this section we generally drop the argument of both $\tpsi(r)$ and $\tilde{H_2}(r)$ for brevity. 

The system~\eqref{eq:deltaPsiH2} consists of the fourth and third order differential equations, which can be combined into the system of two third order differential equations. To see this one has to take a 
coordinate derivative of eq.~\eqref{eq:deltaH2} and then take a linear combination
of the resulting equation and eq.~\eqref{eq:deltaPsi} with a factor
\begin{equation}
\label{eq:constC}
\mathcal{C} = \dfrac{1}{4 \ell(\ell+1) B J } \dfrac{\TT}{\cal{H}},
\end{equation}
so that both 
$(2 g_5 \mathcal{C} + h_6)\tpsi''''$ and $(h_6 \mathcal{C} + 2 f_2)\tilde{H_2}'''$ vanish (see eq.~\eqref{eq:action_coef1}).
Then the resulting system reads as follows:
\begin{subequations}
\label{eq:deltaPsiH2new}
\begin{align}
& [4 g_5' \mathcal{C} + 2 h_6' + h_2]\;\tpsi''' + [2g_5'' \mathcal{C} - 2g_2 \mathcal{C} +h_6'' +2 h_2' - h_3 - \omega^2 (2 g_4 \mathcal{C} +h_5) ]\; \tpsi'' 
+ [ h_2''-2 g_2' \mathcal{C}  - h_3' - h_4 
\nonumber\\\nonumber
&- \omega^2(2 g_4' \mathcal{C} + h_5' + h_1)]\; \tpsi' + [2g_3 \mathcal{C} - h_4' - \omega^2(2 g_1 \mathcal{C} + h_1')] \tpsi
+ [2 h_6' \mathcal{C} - h_2 \mathcal{C} + 4 f_2'] \tilde{H_2}'' 
+ [h_6'' \mathcal{C} - h_2' \mathcal{C} 
\nonumber\\
& 
- h_3 \mathcal{C} + 2f_2'' -2 f_3 - \omega^2 (h_5 \mathcal{C} + 2 f_1)] \tilde{H_2}' - [h_3' \mathcal{C} - h_4 \mathcal{C} + 2 f_3' - \omega^2 (\mathcal{C}(h_5' - h_1) + 2 f_1')] \tilde{H_2} = 0,\\\nonumber
\\
& 2 f_2 \tilde{H_2}'' + 2 f_2 \tilde{H_2}' - 2 f_3 \tilde{H_2} + h_6 \tpsi''' + (h_6'+h_2)\tpsi''
+ (h_2' - h_3)\tpsi' - h_4 \tpsi - \omega^2 ( 2 f_1 \tH + h_5 \tpsi' 
+h_1\tpsi) = 0,
\end{align}
\end{subequations}
where the leading terms with derivative  
are $\tpsi'''$ and $\tilde{H_2}''$.
Let us now prove that the system~\eqref{eq:deltaPsiH2new} admits only four regular solutions in the vicinity of $\TT = 0$.

We arrange the coordinate system so that $\TT$ crosses zero at $r=0$ so we write
\begin{equation}
\label{eq:linear_theta}
\TT = \gamma \cdot r + \dots,
\end{equation}
where $\gamma$ is a regular constant and dots denote higher order 
in $r$. So from now on we focus on the behaviour 
of the system~\eqref{eq:deltaPsiH2new} in the vicinity of $r=0$. 
The coefficients $h_j$, $g_j$ and $f_j$ 
in eqs.~\eqref{eq:deltaPsiH2new} are regular at all points and can be expanded into power-series around $r=0$: 
\begin{equation}
\begin{aligned}
\label{eq:ansatz_hgf}
& h_j = h_{j,0} + h_{j,1} \cdot r + h_{j,2}\cdot r^2  + \dots, \quad (j=\overline{1,4})\\
& h_5 = h_{5,1}\cdot r + h_{5,2}\cdot r^2  + \dots,\\
& h_6 = h_{6,1}\cdot r + h_{6,2}\cdot r^2  + \dots,\\
& g_j = g_{j,0} + g_{j,1}\cdot r + g_{j,2}\cdot r^2  + \dots, \quad (j=\overline{1,5})\\
& f_3 = f_{3,0} + f_{3,1}\cdot r + f_{3,2}\cdot r^2  + \dots,
\end{aligned}
\end{equation}
where $h_{j,k}$, $g_{j,k}$ and $f_{j,k}$ are constants and for 
$f_1$ and $f_2$ we make use of eqs.~\eqref{eq:pefect_squares2} to express them as follows: 
\begin{equation}
\label{eq:f1f2}
f_1 = \frac{h_5^2}{4\; g_4}, \qquad f_2 = \frac{h_6^2}{4\; g_5}.
\end{equation}
Let us note the expansions for $h_5$ and $h_6$ start from linear terms according to their definitions in eqs.~\eqref{eq:coef_tildePsi}.

So upon substitution of 
the expansions~\eqref{eq:linear_theta}-\eqref{eq:f1f2} into 
eqs.~\eqref{eq:deltaPsiH2new}
we find four independent solutions, which to the leading order read:
% we look for solutions $\tpsi(r)$ and $\tilde{H_2}(r)$ at around $r=0$ in the following form:
% \begin{equation}
% \begin{aligned}
% \label{eq:ansatzPsiH2}
% & \tpsi = C_1 r^{\alpha} + C_2 r^{\alpha + 1} + \dots,\\
% & \tilde{H_2} = D_1 r^{\beta} + D_2 r^{\beta + 1} + \dots,
% \end{aligned}
% \end{equation}
% where $C_i$, $D_i$, $\alpha$ and $\beta$ are constants. 
% Let us find the solutions to eqs.~\eqref{eq:deltaPsiH2new} in the leading order of $r$. Substituting eqs.~\eqref{eq:linear_theta},~\eqref{eq:ansatz_hgf},~\eqref{eq:ansatzPsiH2} into eqs.~\eqref{eq:deltaPsiH2new}
% we find four independent solutions:
\begin{subequations}
\label{eq:regular_solutions}
\begin{align}
&\left(
  \begin{array}{c}
    \tpsi \\
    \tilde{H}_2 \\
  \end{array}
\right)
=
c_1\exp(i\omega t)\left(
            \begin{array}{c}
              1 + O(r^4)\\
              \dfrac{-h_{1,0}\cdot\omega^2-h_{4,0}}{2 f_{3,0}}+O(r) \\
            \end{array}
          \right),\\
% \end{equation}
% \begin{equation}
&\left(
  \begin{array}{c}
    \tpsi \\
    \tilde{H}_2 \\
  \end{array}
\right)
=
c_2\exp(i\omega t)\left(
            \begin{array}{c}
              r +O(r^4) \\
              \dfrac{h_{2,1}-h_{3,0}}{2 f_{3,0}} + O(r)\\
            \end{array}
          \right),\\
% \end{equation}
% \begin{equation}
&\left(
  \begin{array}{c}
    \tpsi \\
    \tilde{H}_2 \\
  \end{array}
\right)
=
c_3\exp(i\omega t)\left(
            \begin{array}{c}
              r^2 + O(r^4)\\
              \dfrac{h_{2,0}+h_{6,1}}{f_{3,0}} + O(r)\\
            \end{array}
          \right),\\
% \end{equation}
% \begin{equation}
&\left(
  \begin{array}{c}
    \tpsi \\
    \tilde{H}_2 \\
  \end{array}
\right)
=
c_4\exp(i\omega t)\left(
            \begin{array}{c}
              r^3 + O(r^4)\\
              \dfrac{(6h_{2,0}+12h_{6,1})g_{5,0}}{2g_{5,0}f_{3,0}-h_{6,1}^2} r + O(r^2)\\
            \end{array}
          \right),
\end{align}
\end{subequations}
where $c_i$ are constants.
The solutions~\eqref{eq:regular_solutions} 
are indeed regular at $r=0$, i.e. when $\TT$ crosses zero. 
Let us note that as soon as 
$\tpsi$ and $\tilde{H_2}$ are found the rest of
metric perturbations, namely, $K$, $\alpha$, $H_1$ and $H_0$ can be restored from 
eqs.~\eqref{eq:tildePsi},~\eqref{eq:H0_RW}-\eqref{eq:alpha_shifted} 
and they turn out to be regular as well since denominators in the corresponding relations are non-zero.

By proving that solutions~\eqref{eq:regular_solutions} behave 
regularly around $\TT$-crossing 
we have shown that $\TT=0$ is not a special
point for the action~\eqref{eq:unconstrained_action_tildePsi}. Hence, 
the potential singularities at $\TT=0$ 
in the resulting action for $\bpsi$ and
$\bar{H_2}$ after redefinition~\eqref{eq:tildePsi2} should be 
attributed
to the singularity of the redefinition upon $\TT$-crossing.
Keeping this in mind, 
in the following section we explicitly adopt redefinition~\eqref{eq:tildePsi2}
in order to have the quadratic action for perturbations in a conventional form and 
derive a set of stability conditions for two DOFs $\bpsi$ and $\bar{H_2}$.
% cast the action~\eqref{eq:unconstrained_action_tildePsi} in a similar 
% form to the quadratic action in the old gauge, see eq.~\eqref{eq:even_action_final_old}.

% means that the action~\eqref{eq:unconstrained_action_tildePsi}
% describes two healthy DOFs $\tpsi$ and $\tilde{H_2}$ around $\TT=0$. 

% Hence, it is legitimate to make use of redefinition~\eqref{eq:tildePsi2} keeping in mind that seeming divergencies in the 

%%%%%%%%%%%%%%%%%%%%%%%%%%%%%%%%%%%%%%%%%%%%%%%%%%%%%%%%%%%%%%%%%%%%%%%%
\subsubsection{Stability constraints in Regge-Wheeler-unitary gauge}
\label{sec:new_stability_constraints}

Let us now make use of the redefinition~\eqref{eq:tildePsi2} and cast 
the action~\eqref{eq:unconstrained_action_tildePsi} in a conventional form similar to that in the spherical gauge (cf.~\eqref{eq:even_action_final_old}): 
\begin{equation}
\label{eq:action_KGQM_new}
S_{even}^{(2)} = \int \mbox{d}t\:\mbox{d}r 
% \sqrt{\frac{A}{B}} J^2 
\left(
\frac12 \bar{\mathcal{K}}_{ij} \dot{v}^i \dot{v}^j - \frac12 \bar{\mathcal{G}}_{ij} v^{i\prime} v^{j\prime} - \bar{\mathcal{Q}}_{ij} v^i v^{j\prime} - \frac12 \bar{\mathcal{M}}_{ij} v^i v^j
\right),
\end{equation}
where $i=1,2$ with $v^1 = \bar{H_2}$, $v^2 = \bpsi$ and both DOFs are described by the second order differential equations.
% \marginpar{in WM v1=psi, v2=H2!}
% To check positive definiteness of matrices $\bar{\mathcal{K}}$
% and $\bar{\mathcal{G}}$
The explicit expressions for $\bar{\mathcal{K}}_{11}$, $\det\bar{\mathcal{K}}$, $\bar{\mathcal{G}}_{11}$ and $\det\bar{\mathcal{G}}$ 
are as follows:
{\small
\begin{equation}
\label{eq:k11detk_RW}
\bar{\mathcal{K}}_{11} = 
\dfrac{ 1}{4 \ell(\ell+1)(\ell +2)(\ell-1) \sqrt{AB}}
\dfrac{\TT^2}{\mathcal{F}},
\quad
\det\bar{\mathcal{K}} = \dfrac{\ell(\ell+1) B}{4(\ell+2)(\ell-1)A}
\dfrac{(2\mathcal{H} J J' + \Xi \pi')^2 (2 \mathcal{P}_1 - \mathcal{F})}{\mathcal{F}},
\end{equation}
\begin{equation}
\label{eq:g11detg_RW}
\bar{\mathcal{G}}_{11} = \dfrac{\sqrt{AB} }{4 \ell(\ell+1)(\ell +2)(\ell-1)} \dfrac{\mathcal{H}\TT^2}{\mathcal{F}^2}, \quad
% \end{equation}
% \begin{equation}
\det\bar{\mathcal{G}}=\dfrac{\ell (\ell+1) A B^3}
{ 4(\ell+2) (\ell-1)}
\dfrac{\mathcal{G}  (2 J^2 \Gamma \mathcal{H} \Xi \pi'^2  - \mathcal{G} \Xi^2 \pi'^2-4 J^4 \Sigma \mathcal{H}^2/B )}{ \mathcal{F}^2 },
\end{equation}}
while the rest of components in both matrices are given in Appendix C.
Let us note that $\det\bar{\mathcal{K}}$ differ from its counterpart in the old gauge~\eqref{eq:detKold} by a certain non-negative factor:
\begin{equation}
\label{eq:relationKK}
\det\bar{\mathcal{K}} = \dfrac{\ell^2(\ell+1)^2 AB \pi'^2}{64(\ell-1)^2(\ell+2)^2 J'^2} \dfrac{\cH^2 \TT^2}{\cF^2} \; \det{\mathcal{K}},
\end{equation}
and the same holds for $\det\bar{\mathcal{G}}$ and $\det{\mathcal{G}}$~\eqref{eq:detGold}. The relation~\eqref{eq:relationKK} serves as a cross check for our calculations in the Regge-Wheeler-unitary gauge. 
According to eqs.~\eqref{eq:k11detk_RW} and~\eqref{eq:g11detg_RW}
no-ghost and no radial gradient instabilities constraints are still given by eqs.~\eqref{eq:no_ghost_old} and~\eqref{eq:no_gradient_old}, and sound speeds squared for radial perturbations are also unchanged and are given by eqs.~\eqref{eq:speed_even_old}.

As for potentially divergent coefficients in the action~\eqref{eq:action_KGQM_new} upon $\TT$-crossing
% upon redefinition~\eqref{eq:tildePsi2} 
we see that both
$\det\bar{\mathcal{K}}$, $\det\bar{\mathcal{G}}$, $\bar{\mathcal{K}}_{11}$ and $\bar{\mathcal{G}}_{11}$
are non-singular 
at $\TT=0$ while $\bar{\mathcal{K}}_{22}$ and $\bar{\mathcal{G}}_{22}$ hit singularities (see Appendix C). 
Let us recall that $\TT$ inevitably crosses zero at some point(s)
provided that the no-ghost constraint is satisfied (see Sec.~\ref{sec:old_gauge}).
So in the end we have not avoided 
singular coefficients in the quadratic 
action~\eqref{eq:action_KGQM_new}, but unlike the case of 
action~\eqref{eq:even_action_final_old} in the spherical gauge we have explicitly checked that these divergencies 
appear due to our technical redefinition~\eqref{eq:tildePsi2}
and have no physical nature. 
% while 
% both $\bpsi$ and $\bar{H_2}$ are perfectly regular.
% \marginpar{\bf ???}

The significant difference between the new 
matrices $\bar{\mathcal{K}}_{ij}$, $\bar{\mathcal{G}}_{ij}$ 
and ${\mathcal{K}}_{ij}$, ${\mathcal{G}}_{ij}$
in the old gauge is that the determinants of former ones do not involve $J'$ as a common factor, cf. eqs.~\eqref{eq:k11detk_RW},~\eqref{eq:g11detg_RW} and eqs.~\eqref{eq:detKold},~\eqref{eq:detGold}. 
This fact becomes important in stability analysis for a wormhole
solution, whose characteristic tunnel-like profile is encoded in the metric function
$J(r)$, and $J'(r)=0$ at the narrowest point of the throat. The latter 
made both $\det\mathcal{K}$ and $\det\mathcal{G}$ automatically vanishing, which made the 
stability analysis in the spherical gauge not entirely conclusive close to the center of a wormhole's throat.  

% It immediately follows from eqs.~\eqref{eq:k11detk_RW} and~\eqref{eq:g11detg_RW} that both no-ghost condition~\eqref{eq:no_ghost_old}
% and no radial gradient instabilities constraint~\eqref

Let us now turn to matrices $\bar{\mathcal{Q}}_{ij}$ and $\bar{\mathcal{M}}_{ij}$ in action~\eqref{eq:action_KGQM_new}. We adopt a convention where matrix $\bar{\mathcal{Q}}_{ij}$ has the only non-vanishing element $\bar{\mathcal{Q}}_{12}$, while other elements are included in matrix $\bar{\mathcal{M}}_{ij}$ upon integration by parts.  
In full analogy with the odd-parity sector in Sec.~\ref{sec:odd_sector}
matrices $\bar{\mathcal{Q}}_{ij}$ and 
$\bar{\mathcal{M}}_{ij}$ generally give 
two types of constraints for the linearized theory: one for angular gradient instabilities and the other one for "slow" tachyonic instabilities. In what follows we concentrate on angular gradient instabilities, so it is sufficient for us to consider only those parts  
of both $\bar{\mathcal{Q}}_{ij}$ and $\bar{\mathcal{M}}_{ij}$ which are 
proportional to $\ell(\ell+1)$. 
We have found that in the leading order $\bar{\mathcal{Q}}_{12}$ does
not depend on $\ell$
% does not involve $\ell^2$-part 
so it does not provide any
constraints in the context of angular gradient instabilities, while 
the leading order in angular momentum of both $\bar{\mathcal{M}}^{(\ell^2)}_{11}$ and $\det\bar{\mathcal{M}}^{(\ell^2)}$ read:
% \marginpar{$\ell^2 \to \ell(\ell+1)$?}
\begin{equation}
\label{eq:m11_RW}
\bar{\mathcal{M}}^{(\ell^2)}_{11} = \ell^2  \dfrac{\sqrt{A}}{\sqrt{B}}\dfrac{(\mathcal{H} - 2 F_4 B^2 \pi'^4)^2}{\mathcal{F}},
\end{equation}

{\small
\begin{equation}
\label{eq:detM_RW}
\det\bar{\mathcal{M}}^{(\ell^2)} = - \dfrac{\ell^2 \left( \mathcal{H} - 2 F_4 B^2 \pi'^4\right)^2 \mathcal{H}^2}{16 \mathcal{F}^2 A J^2} 
\left(
8 A^2 \dfrac{\mathcal{F} \mathcal{G}}{\mathcal{H}^2} + B[\mathcal{F}\mathcal{P}_4 - (A'J - 2 A J')]^2 + 4 A^{3/2}B J^4 \mathcal{F} \frac{d}{dr}\left[ \frac{\sqrt{B}}{\sqrt{A} J^3
} \mathcal{P}_4 \right]
\right),
\end{equation}
}
with
\begin{equation}
\mathcal{P}_4 = \dfrac{\mathcal{H} A' J + 2 \mathcal{G} A J' + \Gamma A J \pi'}{\mathcal{H} \left(\mathcal{H} - 2 F_4 B^2 \pi'^4\right)},
\end{equation}
and the rest of matrix components are given in Appendix C.
We have introduced a superscript $\ell^2$ to emphasize that this is only
a part of matrix $\bar{\mathcal{M}}$ that is proportional to $\ell(\ell+1)$.

We see that in full analogy with $\bar{\mathcal{K}}_{ij}$ and $\bar{\mathcal{G}}_{ij}$ matrix $\bar{\mathcal{M}}^{(\ell^2)} _{ij}$ involves 
the key combination $\left( \mathcal{H} - 2 F_4 B^2 \pi'^4\right)$ which comes from $\TT$ (see eq.~\eqref{eq:Theta}) upon taking the leading order in $\ell$ and has to cross zero in order to avoid ghost instabilities, see eq.~\eqref{eq:no_go1}. Hence, the
matrix $\bar{\mathcal{M}}^{(\ell^2)} _{ij}$ behaves in a similar way to 
$\bar{\mathcal{K}}_{ij}$ and $\bar{\mathcal{G}}_{ij}$
upon $\TT$-crossing: $\det\bar{\mathcal{M}}^{(\ell^2)} $ is regular while
$\bar{\mathcal{M}}^{(\ell^2)}_{11}$ crosses zero. 
%and $\bar{\mathcal{M}}_{22}$ diverges.

As before to ensure that angular gradient instabilities are absent 
it is sufficient to require positive definiteness of matrix $\bar{\mathcal{M}}$:
\begin{equation}
\label{eq:positiveM}
\bar{\mathcal{M}}^{(\ell^2)} _{11} > 0, \qquad \det\bar{\mathcal{M}}^{(\ell^2)} >0.
\end{equation}
According to eq.~\eqref{eq:m11_RW} $\bar{\mathcal{M}}^{(\ell^2)} _{11}$ is automatically non-negative, while making the determinant positive
requires the following: 
\begin{equation}
\label{eq:no_ang_grad}
\frac{d}{dr}\left[ \frac{\sqrt{B}}{\sqrt{A} J^3
} \mathcal{P}_4 \right]< - \frac{1}{4 A^{3/2} J^4 }
\left(8 \frac{A^2}{B} \dfrac{ \mathcal{G}}{\mathcal{H}^2} 
+ \frac{[\mathcal{F}\mathcal{P}_4 - (A'J - 2 A J')]^2}{\mathcal{F}} \right).
\end{equation}
Thus, parity even modes have no angular gradient instabilities 
provided that inequality~\eqref{eq:no_ang_grad} is satisfied. The corresponding sound speeds squared propagating along the angular direction $c^2_{a1,2}$ are given as eigenvalues 
of matrix $(A^{-1}J^2)(\bar{\mathcal{K}})^{-1}\bar{\mathcal{M}}^{(\ell^2)} $. We omit their explicit expressions here as they are quite cumbersome and not really illuminating.

We stress again that our stability analysis for the even-parity modes in this section remains incomplete as opposed to the odd-parity modes, see Sec.~\ref{sec:odd_sector}. Namely, we have not addressed 
"slow" tachyonic instabilities, which become significant if we consider all modes, including those with low momentum. As we mentioned above the corresponding stability conditions for tachyons are associated with the remaining parts of matrices $\bar{\mathcal{Q}}_{ij}$ and 
$\bar{\mathcal{M}}_{ij}$ which are lower order in $\ell$. 
Formulating these conditions in a concise form is challenging and
we leave it for future development.

To sum up, the quadratic action~\eqref{eq:action_KGQM_new} gives the following set of stability conditions which ensure the absence of both ghost and gradient instabilities in the even-parity sector
(see eqs.~\eqref{eq:stability_even_ghost},~\eqref{eq:stability_even_radial} and~\eqref{eq:no_ang_grad}):
\begin{subequations}
\label{eq:stability_even}
\begin{align}
\label{eq:stability_K_RW}
\mbox{No ghosts:}\quad &2\frac{\sqrt{B}}{\sqrt{A}} \cdot
\frac{\mbox{d}}{\mbox{d}r}\left[\frac{\sqrt{A}}{\sqrt{B}}
\frac{J^2 \mathcal{H}\left(\mathcal{H} - 2 F_4 B^2 \pi'^4\right)}{2{\cal H} J J' + \Xi \pi'}\right]-\mathcal{F} > 0, \\
\label{eq:stability_G_RW}
\mbox{No radial gradient instabilities:}\quad &2 J^2 \Gamma \mathcal{H} \Xi \pi'^2  - \mathcal{G} \Xi^2 \pi'^2-4 J^4 \Sigma \mathcal{H}^2/B  > 0, \\
\label{eq:stability_M_RW}
\mbox{No angular gradient instabilities:}\quad & \frac{d}{dr}\left[ \frac{\sqrt{B}}{\sqrt{A} J^3
} \mathcal{P}_4 \right]< - \frac{1}{4 A^{3/2} J^4 }
\left(8 \frac{A^2}{B} \dfrac{ \mathcal{G}}{\mathcal{H}^2} 
+ \frac{[\mathcal{F}\mathcal{P}_4 - (A'J - 2 A J')]^2}{\mathcal{F}} \right).
\end{align}
\end{subequations}
So the complete set of stability conditions for high momenta modes in parity odd and parity even sectors  is given by eqs.~\eqref{eq:stability_odd} and~\eqref{eq:stability_even}.

%%%%%%%%%%%%%%%%%%%%%%%%%%%%%%%%%%%%%%%%%%%%%%%%%%%%%%%%%%%%%%%%%%%%%%%
\section{Wormhole beyond Horndeski: an example}
\label{sec:wormhole_example}

In this section we give a specific example of beyond Horndeski Lagrangian~\eqref{eq:lagrangian} which admits a static, spherically-symmetric wormhole solution. Our main requirement to the solution is that it has to be stable against ghosts and gradient instabilities in both parity odd and parity even sectors.
In full analogy with our previous wormhole solution in Ref.~\cite{wormhole1} we adopt a "reconstruction" procedure: we
choose the background metric~\eqref{eq:backgr_metric} of a wormhole form
and concoct the Lagrangian functions so that background equations of motion~\eqref{eq: background_eqs} and stability conditions in eqs.~\eqref{eq:stability_odd} and~\eqref{eq:stability_even}
% ~\eqref{eq:stability_even_ghost},~\eqref{eq:stability_even_radial} and~\eqref{eq:no_ang_grad} 
are satisfied. We also ensure that
the modes in the odd sector propagate at safely subluminal speed~\eqref{eq:sublum_odd}, 
and the same is done for the speeds of the even parity modes propagating in the radial direction~\eqref{eq:speed_even_old}. 
% However, at this point we can only check that the 

We begin with choosing a specific form of metric functions in~\eqref{eq:backgr_metric}:
\[
\label{eq:AJ}
A = 1, \qquad J = \tau\log\left[1+2\cosh\left(\frac{r}{\tau}\right)\right],
\]
where the parameter $\tau \sim R_{min}$
regulates the size of the wormhole
throat at $r=0$. Our choice for the last metric function $B(r)$ is somewhat less straightforward. There is a linear combination of the 
background equations of motion~\eqref{eq: background_eqs} which involves metric functions explicitly, while all the Lagrangian functions get combined into $\mathcal{F}$ and $\cH$ (see eq.~\eqref{eq:cal_FGH}):
\begin{equation}
\label{eq:dY}
A\cdot\mathcal{E}_J + J^2\cdot\mathcal{E}_A = -\frac{\cF}{J^2}+\frac{BA'^2}{4A^2}+\frac{B'J'}{2J}-\frac{A'(JB'+2BJ')}{4AJ}-\frac{BA''}{2A}+B\left(\frac{J'^2}{J^2}+\frac{J''}{J}\right)=0,
\end{equation}
where $\mathcal{E}_J$ and $\mathcal{E}_A$ are given in Appendix A and  we have already set $\cH = 1$ for simplicity (this is a deliberate choice in our model below although not an obligatory one). Stability conditions for the odd parity sector~\eqref{eq:stability_odd} require $\cF > 0$ so
according to eq.~\eqref{eq:dY} we cannot freely choose all
three metric functions. Having fixed $A(r)$ and $J(r)$ in eq.~\eqref{eq:AJ}
we take $B(r)$ as follows:
\begin{equation}
\label{eq:B}
B(r) = 1+ \mbox{sech}\left(\frac{r}{\tau}\right),
\end{equation} 
where $\tau$ is the same parameter as in eq.~\eqref{eq:AJ}. This choice for $B$ ensures that eq.~\eqref{eq:dY} holds for $\cF > \cH$ with already fixed $\cH=1$ above. 

% Hence, we took into account the stability conditions for odd parity modes~\eqref{eq:stability_odd} and also made these modes propagate at safely subluminal speeds~\eqref{eq:speed_odd} (recall that $\cG=\cH$ for quadratic subclass of beyond Horndeski, see eq.~\eqref{eq:cal_FGH}).   

Finally, we choose static, spherically-symmetric scalar field $\pi(r)$, which supports the throat of a wormhole, as follows:
\begin{equation}
\label{eq:pi}
\pi(r)=\tanh\left(\frac{r}{\tau}\right)-1.
\end{equation} 
This completes our arrangement of the background setting.
Our choice for $\pi(r)$, $A$, $B$ and $J$ above has the following asymptotical behaviour as $r \to\pm\infty$:
\begin{equation}
\label{eq:asymp_background}
\pi(r) \to 0, \qquad A(r) = 1, \qquad B(r) \to 1, \qquad J(r) \to r,
\end{equation} 
which means that the space-time
in our set up tends to an empty Minkowski space far away from the wormhole. 

Let us note that we choose the metric functions $A$, $B$ and $J$ above differently as compared to the original set of $A=B=1$ and $J=\sqrt{\tau^2 + r^2}$ in Ref.~\cite{wormhole1}. This is due to the fact that the corresponding equations of motion and stability conditions in the latter case become oversimplified due to random cancellations so that we have less parameters in the model to control stability at the linearized level.

To find the Lagrangian functions for the devised background, 
we choose the
following Ansatz:
\begin{subequations}
\label{ansatz}
\begin{align}
\label{F}
& F(\pi, X) = f_0(\pi) + f_1(\pi)\cdot X + f_2(\pi)\cdot X^2, \\
%& K(\pi, X) = 0, \\
\label{G4}
& G_4(\pi, X) = \frac12 + g_{40}(\pi) + g_{41}(\pi) \cdot X + g_{42}(\pi) \cdot X^2,\\
%& G_5(\pi, X) = 0, \\
\label{F4}
& F_4(\pi, X) = f_{40}(\pi) + f_{41}(\pi) \cdot X,
%&  F_5(\pi, X) = 0,
\end{align}
\end{subequations}
in full analogy with Ref.~\cite{wormhole1}, but here we have an additional term $g_{42}(\pi)$ which will be used to satisfy the new stability condition in the even parity sector~\eqref{eq:no_ang_grad}.

The following steps 
are almost identical to those in Ref.~\cite{wormhole1}, however, all
the formulae become significantly more involved due to non-trivial $J(r)$ and $B(r)$ in eqs.~\eqref{eq:AJ} and~\eqref{eq:B}. 
In Sec.\ref{sec:wormhole_reconstruction} we schematically describe the way we find functions $f_i(\pi)$, $g_{4i}(\pi)$ and $f_{4j}(\pi)$ by making use of stability conditions~\eqref{eq:stability_odd},~\eqref{eq:stability_even} and background equations, but we skip some of the most lengthy intermediate steps of calculations and do not give the resulting Lagrangian functions analytically to avoid bulky expressions.
% In order to avoid bulky expressions we schematically describe the way we find functions $f_i(\pi)$, $g_{4i}(\pi)$ and $f_{4j}(\pi)$ in Sec.\ref{sec:wormhole_reconstruction} but we do not give them in full analytical form and also skip some of the most lengthy intermediate steps of calculations. 
We present all necessary plots in Sec.~\eqref{sec:wormhole_results}  though, which show the graphs of Lagrangian functions we come up with and explicitly prove that the stability conditions are indeed satisfied. The reader that is interested in the results only
may safely skip the following technical section and go straight to Sec.~\eqref{sec:wormhole_results}.

\subsection{Beyond Horndeski Lagrangian functions: reconstruction}
\label{sec:wormhole_reconstruction}

We make use of eq.~\eqref{ansatz} and express $\cH$, $\cF$, 
$\Xi$, $\Sigma$ and $\Gamma$ in terms of $f_i(\pi)$, $g_{4i}(\pi)$ and $f_{4j}(\pi)$:
\begin{subequations}
\begin{align}
\label{eq:D:H}
& {\cal H} = {\cal G} = 1 + 2 g_{40}(\pi) + B \pi'^2 \cdot g_{41}(\pi) +
 \frac12 {B^2} \pi'^4 \cdot [4\; f_{40}(\pi) - 3\; g_{42}(\pi)]  -
 B^3 \pi'^6 \cdot f_{41}(\pi) ,\\
\label{eq:D:F}
& {\cal F} = 1 + 2 g_{40}(\pi) - B  \pi'^2 \cdot g_{41}(\pi) +
 \frac12 B^2  \pi'^4 \cdot g_{42}(\pi), \\
\label{eq:D:Xi}
& \Xi = 4 B J J' \pi' \left[-3 B^2 \pi'^4 \cdot f_{41}(\pi) 
+ 4 B \pi'^2 \cdot f_{40}(\pi)
- 3 B \pi'^2 \cdot g_{42}(\pi) 
+  g_{41}(\pi) \right]\\\nonumber
&\qquad + J^2 \left[ \frac52 B^2 \pi'^4 \cdot g_{42}'(\pi) 
- 3 B \pi'^2 \cdot g_{41}'(\pi)  + 2 \cdot g_{40}'(\pi)\right],\\
\label{eq:D:Sigma}
&\Sigma=- \frac12 B^2 \pi'^2\left( 6 B \left(\frac{A'}{A} + \frac{J'}{J }\right)\frac{J'}{J }  \pi'^2 (5 B  f_{41}(\pi) \pi'^2-4 f_{40}(\pi)+3 g_{42}(\pi)) \right.\\\nonumber&\left.
+ \left(\frac{A'}{A} + 4  \frac{J'}{J }\right)
\pi'(3 g_{41}'(\pi)-5 B  \pi'^2 g_{42}'(\pi)) - 2 g_{41}(\pi) \left(\frac{A'}{A} + \frac{J'}{J }\right) \frac{J'}{J} \right) 
\\\nonumber
& + \frac{1}{2 J^2}B  \pi'^2  \left( 6 B g_{42}(\pi) \pi'^2 + 2 g_{41}(\pi)\right) 
 -\frac{3}{2} B^2 f_{2}(\pi) \pi'^4 -  \frac{1}{2}B  \pi'^2 f_{1}(\pi),\\
\label{eq:D:Gamma}
& \Gamma = \left(2 B  \pi' \frac{A'}{A} + 4 B \pi' \frac{ J'}{ J}\right)
\left[{4 B  \pi'^2} \cdot f_{40}(\pi)  
- {3 B ^2 \pi'^4 \cdot f_{41}(\pi) } + { g_{41}(\pi)} 
- {3 B \pi'^2 \cdot g_{42}(\pi) } \right] \\\nonumber
% & \qquad + 4 B \pi' \frac{ J'}{ J}\left[ {4 B \pi'^2 \cdot f_{40}(\pi)} 
% - {3 B ^2 \pi'^4 f_{41}(\pi) } + { g_{41}(\pi)} 
% -{3 B  \pi'^2 \cdot g_{42}(\pi)} \right] \\\nonumber
& \qquad  + 4 g_{40}'(\pi) - 6 B  \pi'^2 \cdot g_{41}'(\pi) +5 B ^2 \pi'^4 \cdot g_{42}'(\pi),
\end{align}
\end{subequations}
where we do not explicitly substitute the coordinate dependence of $\pi$
from eq.~\eqref{eq:pi} to keep the equations more concise, but we bear in mind that in fact we work with functions of $r$.

First, we recall that we have already fixed $\cH = 1$ (and $\mathcal{G}=1$, see eq.~\eqref{eq:cal_FGH}) for simplicity when we chose $B$ in eq.~\eqref{eq:B}, so now we can make use of eq.~\eqref{eq:D:H} and express $g_{40}(\pi)$ in terms of $f_{40}(\pi)$, $f_{41}(\pi)$, $g_{41}(\pi)$ and $g_{42}(\pi)$:
\begin{equation}
\label{eq:g40}
g_{40}(\pi) = - \frac12 \left[ B \pi'^2 \cdot g_{41}(\pi) +
 \frac12 {B^2} \pi'^4 \cdot [4\; f_{40}(\pi) - 3\; g_{42}(\pi)]  -
 B^3 \pi'^6 \cdot f_{41}(\pi) \right].
\end{equation}
Then, with $A$, $B$ and $J$ set in eqs.~\eqref{eq:AJ},~\eqref{eq:B} 
we can write eq.~\eqref{eq:dY} substituting $\cF$ 
from~\eqref{eq:D:F} and $g_{40}(\pi)$ from eq.~\eqref{eq:g40}
% \begin{equation}
% \begin{aligned}
% \label{eq:dY_ansatz}
% &\frac{1}{ A^2 J^2}[B J^2 A'^2 -
%  A J [2 B A' J' + J (A' B' +  2 B A)] +
%  A^2 (-4  +
%     2  B' J J' +
%     4 B J'^2 +
%     4 B J J'' \\ 
%     & \qquad \qquad \qquad- 8 g_{40}(\pi) +
%     4 B  \pi'^2 \cdot g_{41}(\pi)-
%     2 B^2 \pi'^4 \cdot g_{42}(\pi) )] = 0,
%     \end{aligned}
% \end{equation}
and express $g_{41}$ in terms of $g_{42}(\pi)$, $f_{40}(\pi)$ and
$f_{41}(\pi)$
as follows:
\begin{equation}
\begin{aligned}
\label{eq:g41}
& g_{41}(\pi) = \frac{1}{J^2 } \left(\frac12  B^2 \pi'^4 \cdot f_{41}(\pi) -   {B} \pi'^2
(f_{40}(\pi) -  g_{42}(\pi)) \right) \\
&- \frac{1}{8 \pi'^2 }
\left(\frac{A'^2}{A^2} -
  \frac{[2 B A' J' + J (A' B' +  2 B A)]}{A B J} +
 \frac{2[ B' J J' + 2 B J'^2 + 2 B J J''-2]}{B J^2}\right).
    \end{aligned}
\end{equation}
% Then, with $A$, $B$ and $J$ set in eqs.~\eqref{eq:AJ},~\eqref{eq:B} 
% we can write eq.~\eqref{eq:dY} substituting $\cF$ 
% from~\eqref{eq:D:F}
% \begin{equation}
% \begin{aligned}
% \label{eq:dY_ansatz}
% &\frac{1}{ A^2 J^2}[B J^2 A'^2 -
%  A J [2 B A' J' + J (A' B' +  2 B A)] +
%  A^2 (-4  +
%     2  B' J J' +
%     4 B J'^2 +
%     4 B J J'' \\ 
%     & \qquad \qquad \qquad- 8 g_{40}(\pi) +
%     4 B  \pi'^2 \cdot g_{41}(\pi)-
%     2 B^2 \pi'^4 \cdot g_{42}(\pi) )] = 0,
%     \end{aligned}
% \end{equation}
% where $g_{40}(\pi)$ is already defined in eq.~\eqref{eq:g40}. 
% Taking into account eq.~\eqref{eq:g40} we express $g_{41}$ in terms of $g_{42}(\pi)$, $f_{40}(\pi)$ and
% $f_{41}(\pi)$.
Note that we chose $B(r)$ in eq.~\eqref{eq:B} so that eq.~\eqref{eq:dY} holds for $\cF > \cH$, i.e in our set up $\cF > 1$. Therefore, we have ensured that the stability conditions in the odd-parity sector~\eqref{eq:stability_odd} are satisfied everywhere and both radial and angular speed~\eqref{eq:speed_odd} is luminal at most.

Let us turn to the no-ghost condition in the even parity sector~\eqref{eq:stability_K_RW}:
\[
\label{eq:D:no_go}
\frac{\sqrt{B}}{\sqrt{A}}\cdot \frac{\mbox{d}}{\mbox{d}r}\left[ \frac{\sqrt{A}}{\sqrt{B}}
\frac{J^2 \mathcal{H}\left(\mathcal{H} - 2 F_4 B^2 \pi'^4\right)}{2{\cal H} J J' + \Xi \pi'} \right] > \frac{{\cal F}}{2}.
\]
As we have discussed in Sec.~\ref{sec:old_gauge} to satisfy this no-ghost constraint for all $r$
% inequality~\eqref{eq:D:no_go} 
it is necessary to make 
$\left(\mathcal{H} - 2 F_4 B^2 \pi'^4\right)$ cross zero at some point(s). Since we have fixed $\cH=1$ above, we can choose
\[
\label{eq:D:F4_choice}
2 F_4(\pi,X) B^2 \pi'^4 \equiv 2 \left(f_{40}(\pi) + f_{41}(\pi)\cdot X\right) B^2 \pi'^4 =
w\cdot \mbox{sech}\left(\frac{r}{\tau}+u\right),
\]
so that $\left(\mathcal{H} - 2 F_4 B^2 \pi'^4\right)$ crosses zero twice. 
% The latter in principle makes it possible to have asymptotically 
% vanishing $F_4(\pi,X)$ as $r\to \pm\infty$. 
% \marginpar{right with asymp?} 
We use eq.~\eqref{eq:D:F4_choice} to express $f_{40}(\pi)$ through $f_{41}(\pi)$. In our numerical results below we take specific values of parameters $u$, $w$ and $\tau$ involved in eq.~\eqref{eq:D:F4_choice}, namely, 
\[
u = 0, \quad w=2, \quad \tau =100,
\]
where $\tau$ still defines the size or the wormhole throat. 
We have introduced $u$ in eq.~\eqref{eq:D:F4_choice} to emphasize that it is possible to take $u \neq 0$, which proves that there is no fine-tuning and it is not obligatory to have $F_4'=0$ right at the throat $r=0$. We still take $u=0$ in our calculations for the sake of simplicity.

Next we turn to the denominator of the no-ghost constraint~\eqref{eq:D:no_go}, where $\Xi$ involves yet undefined $g_{42}(\pi)$ and $f_{41}(\pi)$, see eq.~\eqref{eq:D:Xi}. The simplest option that satisfies the stability condition~\eqref{eq:D:no_go} everywhere is to take $\Xi = 0$, so that we can substitute $g_{40}(\pi)$, $g_{41}(\pi)$ and $f_{40}(\pi)$ from eqs.~\eqref{eq:g40},~\eqref{eq:g41},~\eqref{eq:D:F4_choice} and find $f_{41}(\pi)$ in terms of $g_{42}(\pi)$ from the
the resulting equation.   

The constraint that governs angular gradient instabilities in the even
parity sector~\eqref{eq:stability_M_RW} involves the only yet unconstrained coefficient
$\Gamma$. To satisfy inequality~\eqref{eq:stability_M_RW} everywhere it turned out to be sufficient to take $\Gamma = 0$, so we make use of eq.~\eqref{eq:D:Gamma} and
express $g_{42}(\pi)$ as it is the only function that has not been defined above. 
Thus, at this step we have a closed system for $g_{40}(\pi)$, $g_{41}(\pi)$, $g_{42}(\pi)$, $f_{40}(\pi)$ and $f_{41}(\pi)$ so this 
completes the definition of $G_4(\pi,X)$
and $F_4(\pi,X)$ in eqs.~\eqref{G4} and~\eqref{F4}.

In full analogy with Ref.~\cite{wormhole1} 
% the rest of 
% functions in the Ansatz~\eqref{ansatz}, 
% i.e. $f_0(\pi)$, $f_1(\pi)$
% and $f_2(\pi)$ are to be found 
we find $f_0(\pi)$ and $f_1(\pi)$
from solving two background equations of 
motion~\eqref{eq: background_eqs}, e.g. $\mathcal{E}_A=0$ and $\mathcal{E}_B=0$ (recall that we have already used a linear combination of $\mathcal{E}_A$ and $\mathcal{E}_J$ in eq.~\eqref{eq:dY}, while $\mathcal{E}_{\pi}$ is linearly dependent on the other three equations). As before the final function $f_2(\pi)$ is used to satisfy the stability condition against
radial gradient instabilities~\eqref{eq:stability_G_RW} in parity even sector and simultaneously ensure that the speed $c_{s2}^2 \leq 1$ in eq.~\eqref{eq:speed_even_old} ($c^2_{s1}$ is already set to be subluminal above). Both requirements can be written in a compact form:
\[
\label{D:constraint_f2}
0 < 2 J^2 \Gamma \mathcal{H} \Xi \pi'^2  - \mathcal{G} \Xi^2 \pi'^2-4 J^4 \Sigma \mathcal{H}^2/B \leq (2 {\cal H} J J' + \Xi \pi' )^2 (2 {\cal P}_1 -{\cal F}),
\]
where $\Sigma$ involves $f_2(\pi)$, see eq.~\eqref{eq:D:Sigma}. 
The combination on the right hand side coincides with the no-ghost condition~\eqref{eq:stability_K_RW} and we have already insured its positivity, so we choose $\Sigma$ appropriately to have $c_{s2}^2 \leq 1$:
\[
\label{eq:sigma_choice}
c^2_{s2} = \frac{(2 J^2 \Gamma \mathcal{H} \Xi \pi'^2  - \mathcal{G} \Xi^2 \pi'^2-4 J^4 \Sigma \mathcal{H}^2/B)}{(2 {\cal H} J J' + \Xi \pi' )^2 (2 {\cal P}_1 -{\cal F})} = 
\frac{\cosh\left(\frac{r}{\tau}\right)}{1+\cosh\left(\frac{r}{\tau}\right)},
\]
and express $f_2(\pi)$ by making use of eq.~\eqref{eq:D:Sigma} and recalling that we fixed $\Xi=0$ in our model. This completes the reconstruction of the function $F(\pi,X)$.

\subsection{Beyond Horndeski Lagrangian functions: results}
\label{sec:wormhole_results}

\begin{figure}[H]\begin{center}\hspace{-1cm}
{\includegraphics[width=0.5\textwidth]{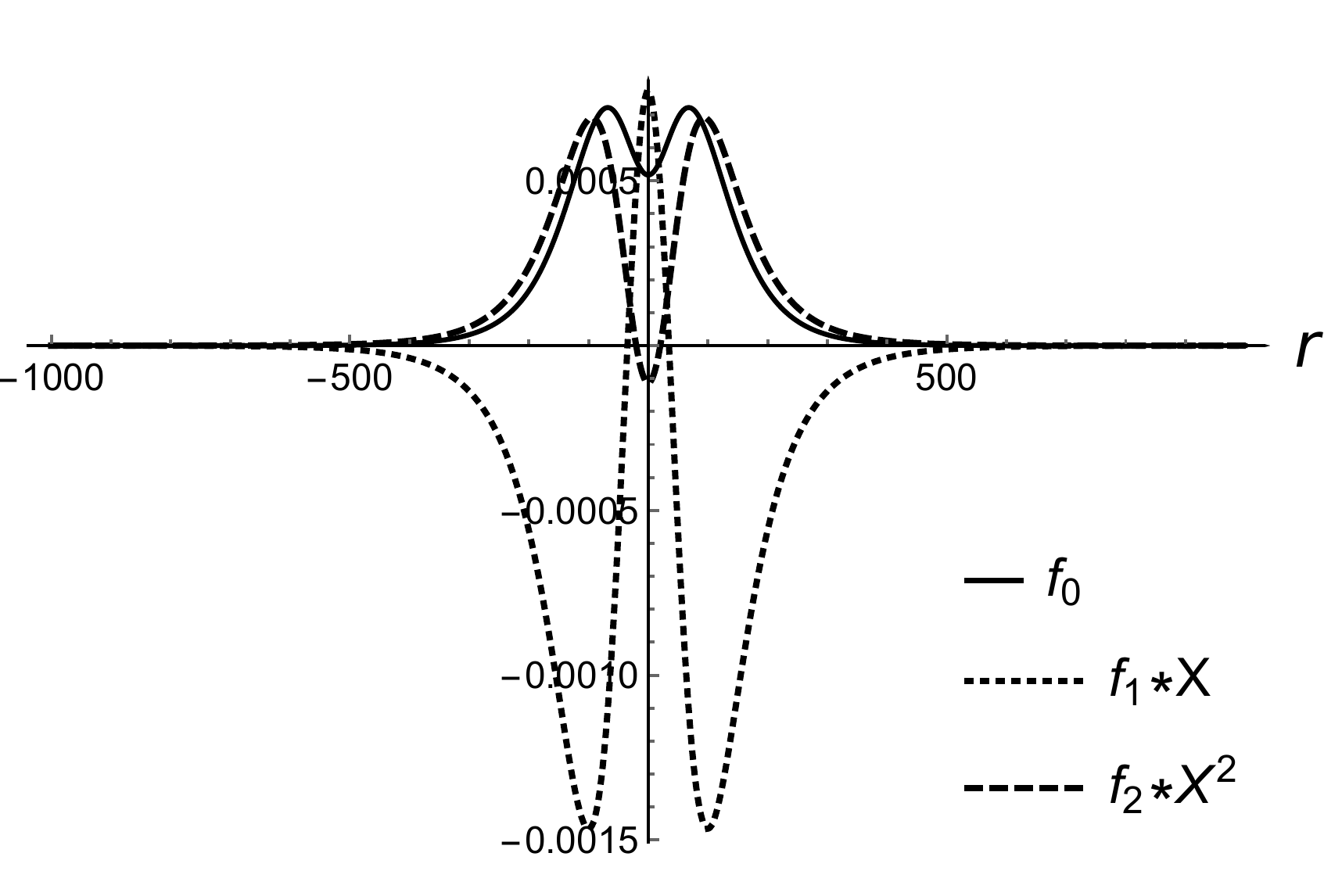}}\hspace{2.8cm}\hspace{-3cm}
{\includegraphics[width=0.5\textwidth] {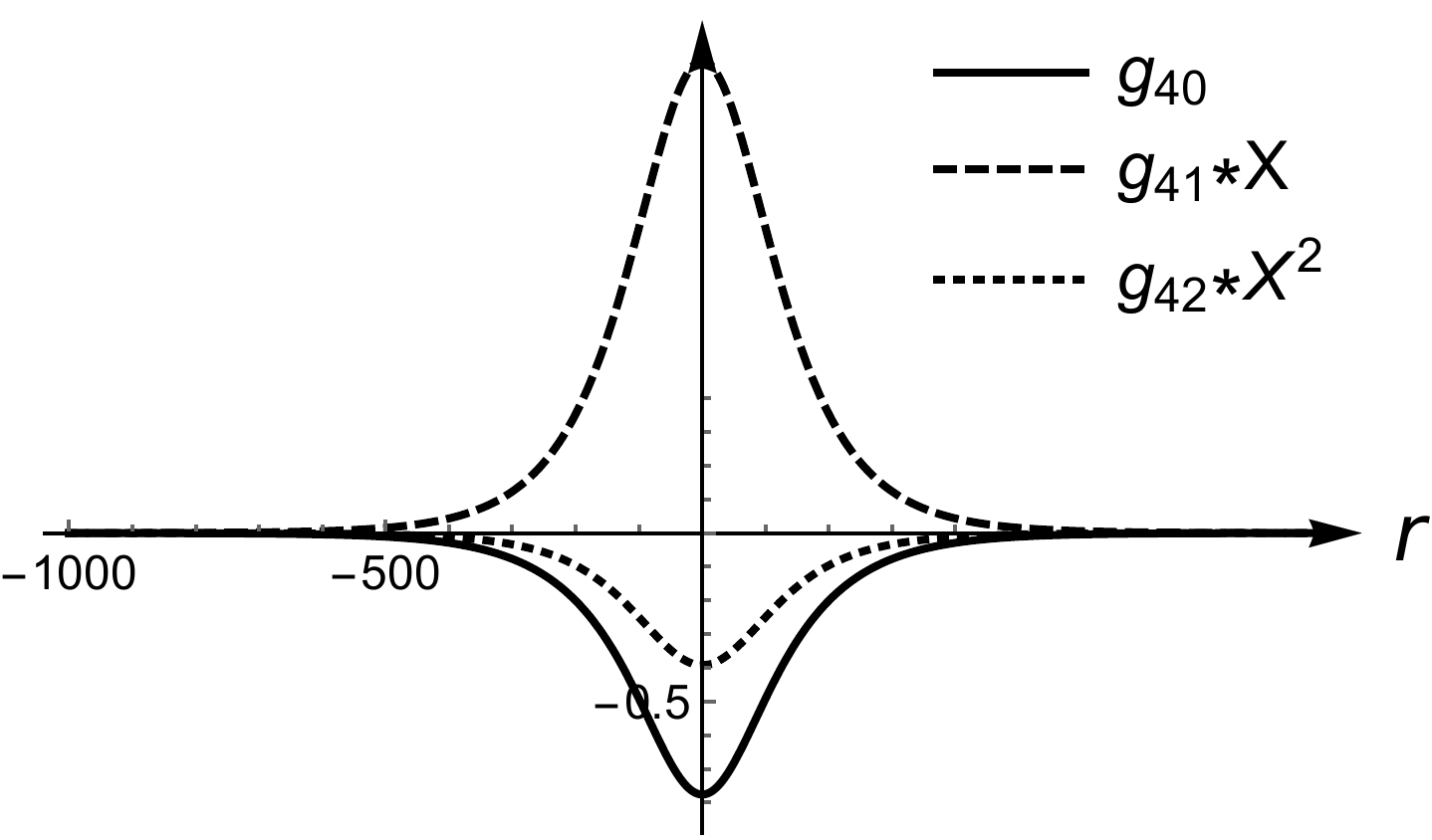}}

\vspace{-0.2cm}
{\includegraphics[width=0.5\textwidth]
{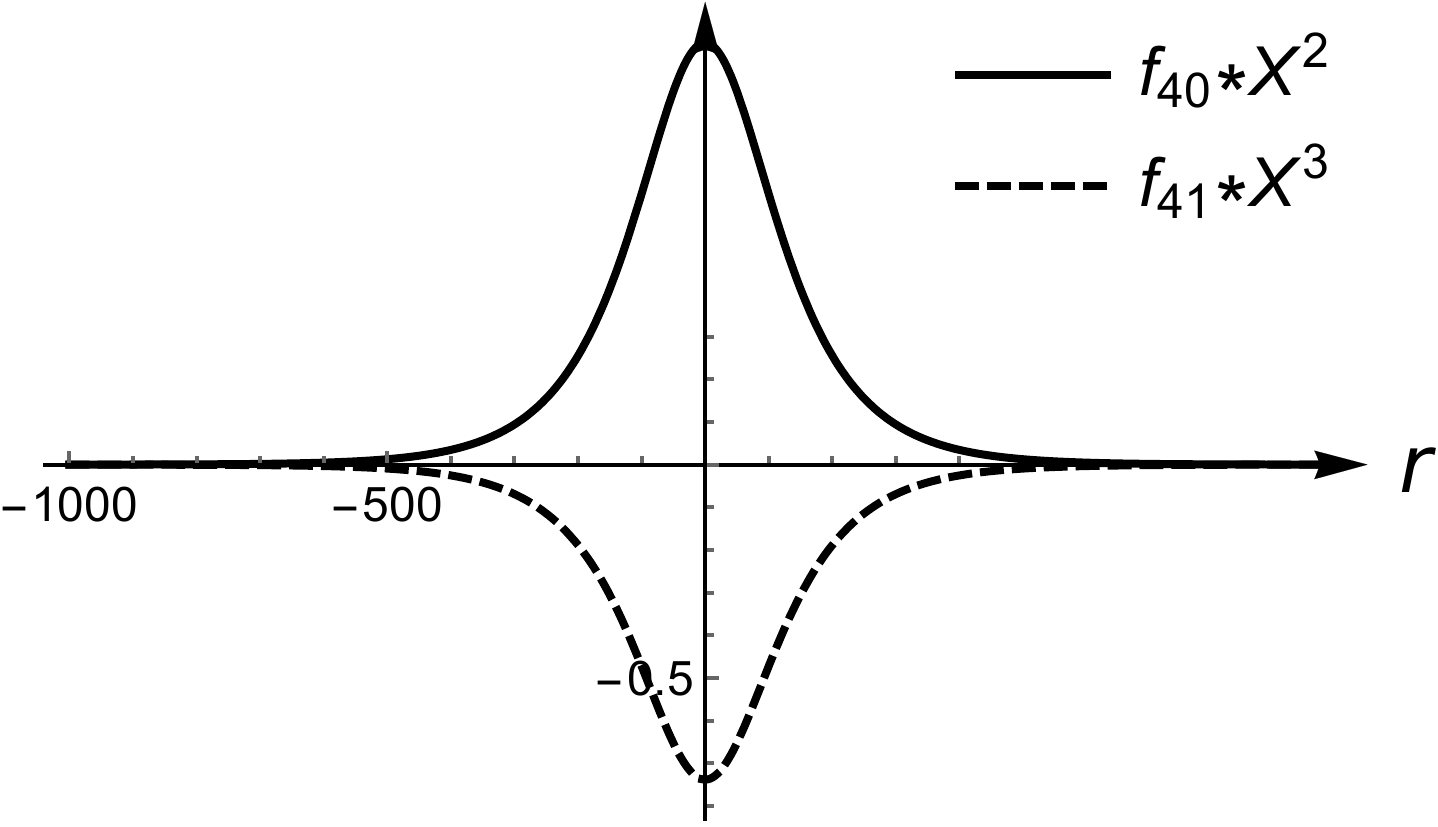}}\hspace{1cm}
\caption{\footnotesize{The Lagrangian functions $f_0(r)$, $f_1(r) \cdot X$, $f_2(r)\cdot X^2$, $g_{40}(r)$, $g_{41}(r)\cdot X$, 
$g_{42}(r)\cdot X^2$,
    $f_{40}(r)$ and $f_{41}(r) \cdot X $, with the following choice of the parameters: $u=0$, $w=2$ and $\tau = 100$. This choice
    guarantees that the size
    of the wormhole throat safely exceeds the Planck length.}} \label{fig:f-g-f4}
\end{center}\end{figure}

The reconstructed functions $f_i(\pi)$, $g_{4i}(\pi)$ and $f_{4j}(\pi)$
entering eqs.~\eqref{ansatz} are given in Fig.~\ref{fig:f-g-f4} and have the following asymptotic behaviour as $r \to \pm \infty$:
\[
\label{eq:fg4f4_asymp}
f_{0}\propto e^{-\frac{r}{\tau}},\quad 
f_{1} \propto e^{\frac{3r}{\tau}}, \quad
f_{2}\propto e^{\frac{7r}{\tau}}, \quad
g_{40} \propto e^{-\frac{r}{\tau}}, \quad
g_{41} \propto e^{\frac{3r}{\tau}}, \quad 
g_{42} \propto e^{\frac{7r}{\tau}}, \quad
f_{40}  \propto e^{\frac{7r}{\tau}},\quad 
f_{41}  \propto e^{\frac{11 r}{\tau}} \; ,
\] 
or, equivalently,
\begin{equation}
\begin{aligned}
\label{eq:fg4f4XXX_asymp}
&f_{0} =\frac{3}{\tau^2}e^{-\frac{r}{\tau}},\quad 
f_{1} \cdot X = -\frac{6}{\tau^2}e^{-\frac{r}{\tau}}, \quad
f_{2} \cdot X^2 =\frac{1}{\tau^2}e^{-\frac{r}{\tau}}, \quad
g_{40} = -\frac{13}{8}e^{-\frac{r}{\tau}}, \quad
g_{41}\cdot X = \frac{9}{4}e^{-\frac{r}{\tau}}, \\
&\qquad\qquad 
g_{42} \cdot X^2 = -\frac{5}{8}e^{-\frac{r}{\tau}}, \quad
f_{40}  = \frac{15}{8}e^{-\frac{r}{\tau}},\quad 
f_{41} \cdot X = -\frac{11}{8}e^{-\frac{r}{\tau}} \; ,
\end{aligned}
\end{equation}
which shows that all terms equally contribute to the Lagrangian functions. Thus, the Lagrangian functions $F(\pi,X)$, $G_4(\pi,X)$
and $F_4(\pi,X)$
have the following asymptotics far away from the throat (recall that $\pi(r)$ in eq.~\eqref{eq:pi})
\[
\label{eq:FG4F4asymp}
\begin{aligned}
&F(\pi, X) = q_1\cdot(-\pi)^{1/2} + q_2\cdot (-\pi)^{-3/2}\cdot X + q_3\cdot (-\pi)^{-7/2}\cdot X^2, \\
&G_4(\pi, X) = \frac12 + q_4\cdot(-\pi)^{1/2} +  q_5\cdot (-\pi)^{-3/2}\cdot X + q_6 \cdot (-\pi)^{-7/2}\cdot X^2, \\
&F_4(\pi, X) = q_7 \cdot (-\pi)^{-7/2} + q_8 \cdot (-\pi)^{-11/2} \cdot X,
\end{aligned}
\]
where $q_i$, $i = \overline{1,8}$ are constants, which are irrelevant at this point. 
% We see that even far away from the throat the theory is still of beyond Horndeski type with significantly modified gravity. 
We see that all functions in eq.~\eqref{eq:fg4f4XXX_asymp} decay 
exponentially with growing $r$, which means that, 
while our theory is still of beyond Horndeski type away from the 
wormhole, the space-time quite rapidly becomes empty Minkowski space
with pure Einstein gravity. 
% as compared to power-laws for $f_i$
% and $f_{4i}$ in our previous solution in Ref.~\cite{wormhole1}

% even far away from the throat the theory is still of beyond Horndeski type with significantly modified gravity. 

%%%%%%%%%%%%%%%%%%%%%%%%%%%%%%%%%%%%%%%%%%%%%%%%%%%%%%%%%%%%%%%%%%%%%%%%
%%%%%%%%%%%%%%%%%%%%%%%%%%%%%%%%%%%%%%%%%%%%%%%%%%%%%%%%%%%%%%%%%%%%%%%%

\begin{figure}[H]\begin{center}
{\includegraphics[width=0.49\textwidth]
{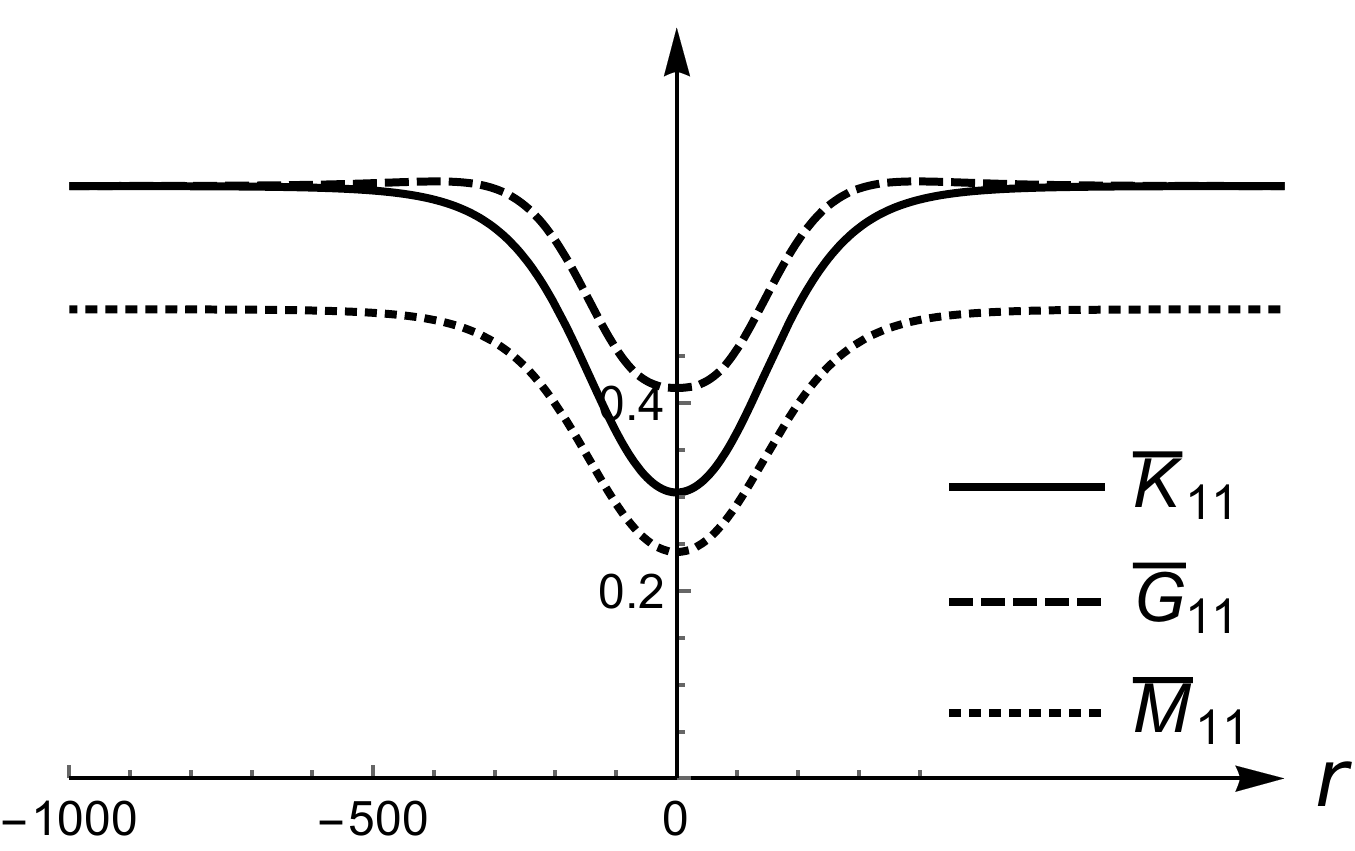}}
\hspace{0.01cm}
{\includegraphics[width=0.49\textwidth]
{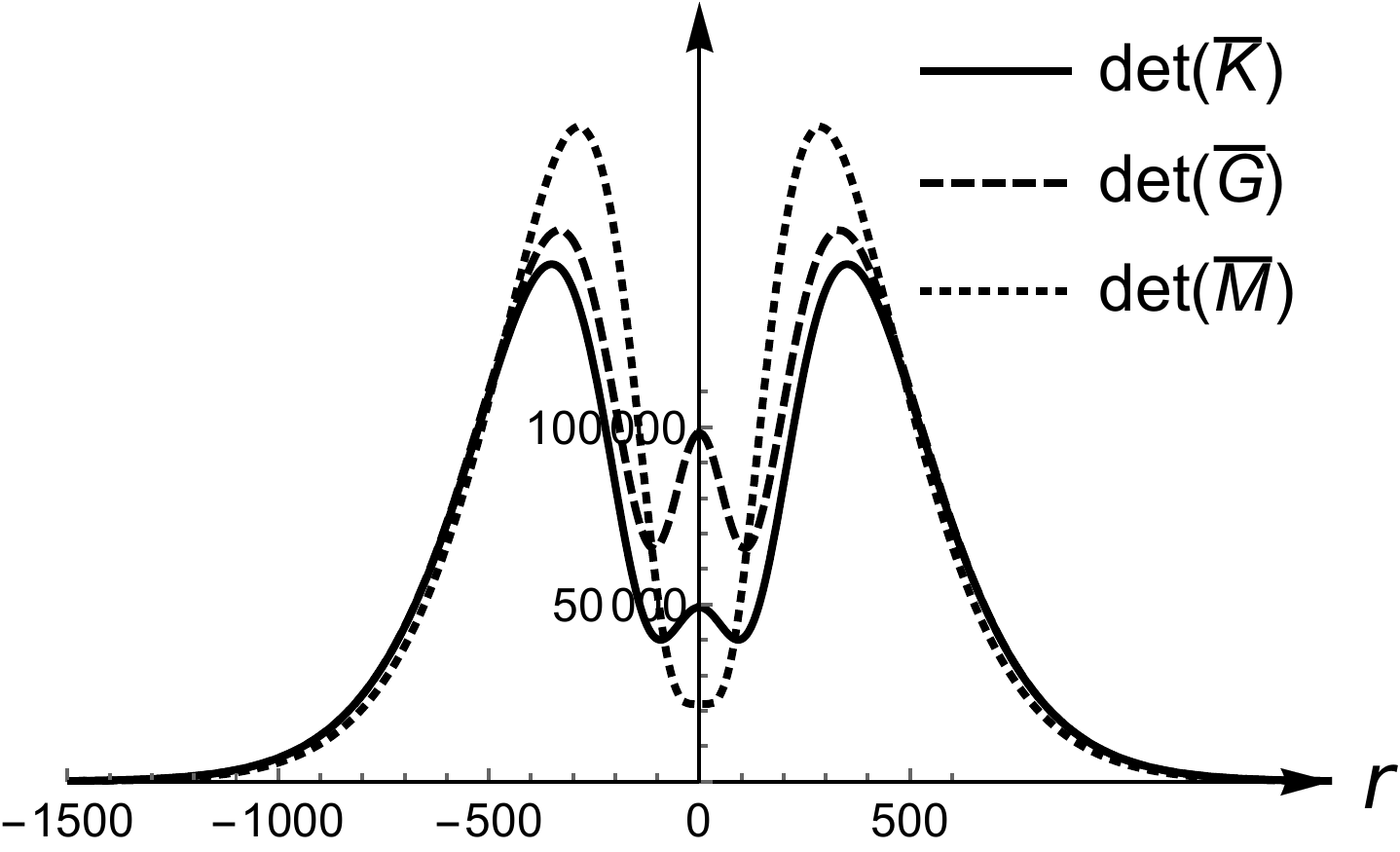}}
\caption{\footnotesize{Functions $\bar{\mathcal{K}}_{11}$, 
$\det\bar{\mathcal{K}}$, $\bar{\mathcal{G}}_{11}$, $\det\bar{\mathcal{G}}$, $\bar{\mathcal{M}}^{(\ell^2)} _{11}$ and $\det\bar{\mathcal{M}}^{(\ell^2)} $, which govern the stability of the 
parity even sector (we choose $\ell = 10$ here for definiteness). We have introduced a normalizing numerical factor for both $\bar{\mathcal{M}}^{(\ell^2)} _{11}$ and $\det\bar{\mathcal{M}}^{(\ell^2)}$, which enabled us to put them on the same plots with matrices $\bar{\mathcal{K}}$ and $\bar{\mathcal{G}}$.
}}
\label{fig:KGM}
\end{center}\end{figure}

Let us demonstrate that our wormhole solution
~\eqref{eq:AJ},~\eqref{eq:B}-\eqref{eq:pi} 
satisfies our set of stability conditions. We have arranged the Lagrangian functions so that $\mathcal{H}=\mathcal{G}=1$ and 
$\mathcal{F} > 1$ which complies with the constraints in the
odd-parity sector~\eqref{eq:stability_odd},~\eqref{eq:sublum_odd} 
and gives at most luminal sound speed squared, i.e. $c^2_{\theta}=1$ and $c^2_r < 1$~\eqref{eq:speed_odd}. As for the parity even modes we show in Fig.~\ref{fig:KGM} that for our set of Lagrangian functions 
$\bar{\mathcal{K}}_{11}$, 
$\det\bar{\mathcal{K}}$, $\bar{\mathcal{G}}_{11}$, $\det\bar{\mathcal{G}}$, $\bar{\mathcal{M}}^{(\ell^2)} _{11}$, $\det\bar{\mathcal{M}}^{(\ell^2)} $ are positive everywhere. Hence, the constraints
% ~\eqref{eq:no_ghost_old},~\eqref{eq:no_gradient_old} and~\eqref{eq:positiveM}
~\eqref{eq:stability_even} 
are satisfied at all points.

%%%%%%%%%%%%%%%%%%%%%%%%%%%%%%%%%%%%%%%%%%%%%%%%%%%%%%%%%%%%%%%%%%%%%%%%
%%%%%%%%%%%%%%%%%%%%%%%%%%%%%%%%%%%%%%%%%%%%%%%%%%%%%%%%%%%%%%%%%%%%%%%%

As for the sound speeds squared in the even-parity sector~\eqref{eq:speed_even_old} 
$c^2_{s1} = c^2_r < 1$, while $c^2_{s2}$ was fixed in eq.~\eqref{eq:sigma_choice} and obviously does not exceed unity. 
The situation, however, is no so bright for the angular sound speeds squared $c^2_{a1,2}$, which are given by eigenvalues of matrix 
$(A^{-1}J^2)(\bar{\mathcal{K}})^{-1}\bar{\mathcal{M}}^{(\ell^2)}$. Even though we do not give these speeds explicitly we can do a quick check if their product is greater than unity, which would signal superluminal propagation in the angular direction. Indeed, the product of $c^2_{a1}$ and $c^2_{a2}$
is given by
\[
c^2_{a1} \cdot c^2_{a2} = \frac{J^4}{A^2}\frac{\det\bar{\mathcal{M}}^{(\ell^2)}}{\det\bar{\mathcal{K}}},
\]
where $\det\bar{\mathcal{K}}$ and $\det\bar{\mathcal{M}}^{(\ell^2)}$ are given in eqs.~\eqref{eq:k11detk_RW} and~\eqref{eq:detM_RW}, respectively.
Our numerical check has shown that for our choice of Lagrangian functions this ratio is greater than one, meaning that either of speeds
$c^2_{a1}$ or $c^2_{a2}$ (or both) is greater than the speed of light. This is another troublesome 
drawback of our solution and we hope to deal with it in
future studies.

\vspace{-0.2cm}
%%%%%%%%%%%%%%%%%%%%%%%%%%%%%%%%%%%%%%%%%%%%%%%%%%%%%%%%%%%%%%%%%%%%%%%%
\section{Conclusion}
\label{sec:outlook}
To sum up, this work aimed to support the point that scalar-tensor theories of modified gravity like beyond Horndeski theories are promising candidates admitting stable Lorentzian wormholes.   
We have formulated a set of stability conditions which eliminate 
high energy pathologies like ghosts and gradient instabilities.
Although imperfect we have put forward a wormhole solution 
that complies with this set of stability constraints.
Unfortunately, we still cannot say anything definite about the tachyon
sector, since dealing with mass matrices is technically demanding
in this class of field theories. It remains a challenging and intriguing
question whether it is possible to have a completely stable wormhole in theories like generalised Horndeski family, let alone make it realistic and observationally traceable. 

\vspace{-0.2cm}
%%%%%%%%%%%%%%%%%%%%%%%%%%%%%%%%%%%%%%%%%%%%%%%%%%%%%%%%%%%%%%%%%%%%%%
%%%%%%%%%%%%%%%%%%%%%%%%%%%%%%%%%%%%%%%%%%%%%%%%%%%%%%%%%%%%%%%%%%%%%%
 \section*{Acknowledgements}{}

When this work was close to completion, Valery Rubakov untimely deceased. All the findings in this paper are the result of our fruitful and long-lasting collaboration with Professor Rubakov. Unfortunately, he did not take part in preparing this manuscript.
\\ 
\\
The work of S.M. and V.V. 
on Sec.~\ref{sec:linearized_th} of this paper has been
supported by Russian Science Foundation grant
19-12-00393, while the part of work on Sec.~\ref{sec:wormhole_example}
has been supported by the Foundation for the Advancement of Theoretical Physics and Mathematics “BASIS”.

\section*{Appendix A}
\label{app:A}
In this Appendix we give Einstein and scalar field equations of motion for action~\eqref{eq:lagrangian} in the background~\eqref{eq:backgr_metric}:
\[
\mathcal{E}_A = 0, \qquad \mathcal{E}_B = 0,
\qquad \mathcal{E}_J = 0, \qquad \mathcal{E}_{\pi} = 0,
\]
where
{\small
\begin{eqnarray}
\hspace{-1cm}
&& \label{CalEa} {\cal E}_A =  F + \frac{2}{J}\left(\frac{1 - BJ'^2}{J} - (2BJ'' + J'B')\right)G_{4}
%\nonumber \\&&
 +\frac{4BJ'}{J}\left( \frac{J'}{J } + \frac{ X'}{X} 
% \right.\nonumber \\&&\left.
 + \frac{2BJ'' + J'B'}{BJ'}\right)XG_{4X} 
\nonumber\\&&
 + \frac{8BJ'}{J}XX'G_{4XX} -
    B\pi'\left(\frac{4J'}{J } + \frac{ X'}{X} \right)G_{4\pi}+ 2B\pi'\left(\frac{4J'}{J } - \frac{ X'}{X} \right)XG_{4\pi X}+ 4XG_{4\pi\pi}
\nonumber\\&&
    -\frac{2B^2\pi'^3}{J^2}(5B'JJ' + B(J'^2 + 2JJ''))\pi' F_{4}
       -\frac{16 B^3 J' \pi'^3}{J} \pi'' F_4 
      + \frac{2B^3J'\pi'^5}{J} \left(B'\pi' + 2B\pi''\right)F_{4X}
      \nonumber\\&&
       - \frac{4B^3J'\pi'^5}{J}F_{4\pi} , \\\nonumber
\end{eqnarray}}
{\small
\begin{eqnarray}
%%
%\\
\nonumber
&&\label{CalEb} {\cal E}_B = F - 2XF_{X}  + \frac{2}{J}\left(\frac{1 - BJ'^2}{J } - BJ'\frac{A'}{A} \right)G_{4}
 -\frac{4}{J}\left(\frac{1 - 2BJ'^2}{J } - 2BJ'\frac{A'}{A}\right)XG_{4X}
- \left(\frac{4J'}{J } + \frac{ A'}{A} \right)B\pi'G_{4\pi}
 \nonumber\\&& 
 + 8\frac{BJ'}{J}\left( \frac{J'}{J } + \frac{ A'}{A}\right)X^2G_{4XX} 
    - 2\left(\frac{4J'}{J } + \frac{ A'}{A} \right)B\pi'XG_{4\pi X} 
     -\frac{10B^3J'(A'J + AJ')\pi'^4}{AJ^2} F_{4}
     \nonumber\\&&
    +\frac{2B^4J'(A'J + AJ')\pi'^6}{AJ^2} F_{4X},\\\nonumber
\end{eqnarray}}
{\small
\begin{eqnarray}
&&\label{CalEc} {\cal E}_J = F 
% \nonumber\\&&
    -\left(\frac{1}{J}\frac{\sqrt{B}}{\sqrt{A}}\left(J\frac{\sqrt{B}}{\sqrt{A}}A'\right)' + \frac{2BJ'' + J'B'}{J} \right)G_{4} - B\pi'\left(\frac{2J'}{J } + \frac{ A'}{A } + \frac{2BJ'' + J'B'}{BJ'} +
      2\frac{\pi'' - \frac{J''}{J'}\pi'}{\pi'}\right)G_{4\pi} 
\nonumber\\&&
  +  BX\left(-\frac{A'^2}{A^2} + \frac{2}{J}\frac{2BJ'' + J'B'}{B} + \frac{A'}{A}\frac{2BJ'' + J'B'}{BJ'} + 2\frac{A'J' + J(A'' - \frac{J''}{J'}A')}{JA}\right)G_{4X} 
\nonumber\\&&
    + BX'\left(\frac{2J'}{J } + \frac{ A'}{A} \right)G_{4X} + 4XG_{4\pi \pi} +
    2B\pi'\left(\frac{2J'}{J } + \frac{ A'}{A } - \frac{ X'}{X}\right)XG_{4\pi X} + 2B\left(\frac{2J'}{J } + \frac{ A'}{A} \right)XX'G_{4XX}
\nonumber\\&&
-\frac{B^2\pi'^4}{2}\left(-\frac{A'^2B}{A^2} + \frac{A'(5B'J + 2BJ')}{AJ} + 2\frac{A''BJ + 5AB'J' + 2ABJ''}{AJ}\right) F_{4}
\nonumber\\&&
 - \frac{B^3(A'J + 2AJ')}{AJ} \pi'^3 \left( 4 \pi'' F_{4} + \pi'^2 F_{4\pi} -
    \frac{\pi'^2(B'\pi' + 2B\pi'')}{2} F_{4X} \right),
    %%   \\
%%
\end{eqnarray}}
and
\begin{eqnarray}
&&\label{CalEphi} {\cal E}_{\pi} = \frac{1}{J^2}\sqrt{\frac{B}{A}} \cdot
\frac{\mbox{d}}{\mbox{d}r}\left[J^2 J' \sqrt{AB} \cdot
\mathcal{J}_H\right] - \mathcal{S}_H + \mathcal{JS}_{BH},
\end{eqnarray}
with
\begin{eqnarray}
&&\mathcal{J}_{H} = \frac{\pi'}{J'}F_{X} + 2\frac{\pi'}{J'}\left(\frac{1 - BJ'^2}{J^2 } - \frac{BJ'}{J}\frac{A'}{A}\right)G_{4X} 
 - \frac{4B\pi'}{J}\left( \frac{J'}{J } + \frac{ A'}{A}\right)XG_{4XX}
 \nonumber \\&&
    - 2\left( \frac{4}{J } + \frac{ A'}{AJ'}\right)XG_{4\pi X},\\\nonumber
    \\
% \end{eqnarray}
% \begin{eqnarray}
  &&
\mathcal{S}_{H} = -F_{\pi} -
 \left[\frac{B}{2}\left( \frac{A'}{A} \right)^2 + \frac{2}{J}\left(\frac{1 - BJ'^2}{J } - (2BJ'' + J'B')\right)\right]G_{4\pi}  
 + B\left[\frac{X'}{X}\left(\frac{4J'}{J } + \frac{ A'}{A} \right) 
\right.\nonumber \\&&\left.
 + \frac{4J'}{J}\left( \frac{J'}{J } + \frac{ A'}{A}\right)\right]XG_{4\pi X}
 % \nonumber\\
 % &&
+
      \left[\frac{B}{2}\left(2\frac{A'' - \frac{J''}{J'}A'}{A} + \frac{A'}{A}\left(\frac{4J'}{J } + \frac{2BJ'' + J'B'}{BJ'}\right)\right)\right]G_{4\pi}
% \nonumber \\&&
%        + B\left[\frac{X'}{X}\left(\frac{4J'}{J } + \frac{ A'}{A} \right) + \frac{4J'}{J}\left( \frac{J'}{J } + \frac{ A'}{A}\right)\right]XG_{4\pi X},
\end{eqnarray}
\begin{eqnarray}
     &&
\mathcal{JS}_{BH} = -4B^2\pi'^2\left[\frac{-A'^2BJ'}{A^2J} \pi'+ \frac{J'(2A''BJ + 5AB'J' + 4ABJ'')}{AJ^2}\pi'
\right.
\nonumber\\ &&
\left.
 + \frac{A'(5B'JJ' + 3BJ'^2 + 2BJJ'')}{AJ^2} \pi' +
        \frac{6BJ'(A'J + AJ')}{AJ^2}\pi''\right]F_{4}
\nonumber\\&&
 -\frac{6B^3J'(A'J + AJ')\pi'^4}{AJ^2}F_{4\pi} +
    B^3\pi'^4\left[-\frac{A'^2BJ'}{A^2J}\pi' + \frac{J'(2A''BJ + 11AB'J' + 4ABJ'')}{AJ^2}\pi'
\right.
\nonumber\\&&
     \left.
    + \frac{A'(11B'JJ' + 3BJ'^2 + 2BJJ'')}{AJ^2}
       \pi' +  \frac{18BJ'(A'J + AJ')}{AJ^2}\pi''\right]F_{4X}
       \nonumber\\&&
 + \frac{2B^4J'(A'J + AJ')\pi'^6}{AJ^2}F_{4 \pi X} -
    \frac{B^4J'(A'J + AJ')\pi'^6(B'\pi' + 2B\pi'')}{AJ^2}F_{4XX}.
\end{eqnarray}

%%%%%%%%%%%%%%%%%%%%%%%%%%%%%%%%%%%%%%%%%%%%%%%%%%%%%%%%%%%%%%%%%%%%%%%%

\section*{Appendix B}
In this Appendix we gather the explicit form of coefficients
$a_i$, $b_i$, $c_i$, $d_i$, $e_i$ and $p_i$ entering the quadratic action in the spherical gauge~\eqref{eq:action_even} and 
its counterpart in
Regge-Wheeler-unitary gauge~\eqref{eq:action_even_RW}. The coefficients 
are given in terms of combinations $\cH$, $\cF$, $\cG$, $\Xi$, $\Gamma$
and $\Sigma$ introduced in the main body of the text in eqs.~\eqref{eq:cal_FGH},~\eqref{eq:Xi},~\eqref{eq:Gamma} and~\eqref{eq:KSI} respectively:
\begin{eqnarray}%{}
a_1&=&\sqrt{AB}\,\Xi,\\
a_2&=&\frac{\sqrt{AB}}{2\pi'}\left[
2\pi'\Xi'-\left(2\pi''-\frac{A'}{A}\pi'\right)\Xi
+2JJ'\left(\frac{A'}{A}-\frac{B'}{B}\right){\cal H}-4{\cal H}JJ''
\right.\\ &&\left.
+\frac{2J^2}{B}\left({\cal E}_B-{\cal E}_A\right) \right] - 
 2  \sqrt{AB^3} \pi'^3 \cdot F_4 ,\\
a_3&=&-\frac{\sqrt{AB}}{2}\left(\pi'\Xi+2JJ'{\cal H}\right),\\
a_4&=&\sqrt{AB}\,{\cal H},\\
a_5&=&-\sqrt{\frac{A}{B}}J^2\frac{\partial{\cal E}_A}{\partial\pi }=a_2'-a_1'',\\
a_6&=&-\sqrt{\frac{A}{B}} \frac{1}{J\pi'} \left( J {\cal H}'+J'{\cal H}-J'{\cal F} \right) + \sqrt{AB} \pi'^2 \left( B' \pi' + 2 B \pi''\right) \cdot F_4, \\
a_7&=&a_3'+\frac{J^2}{2} \sqrt{\frac{A}{B}} {\cal E}_B,
\end{eqnarray}
\begin{eqnarray}
a_8&=&-\frac{a_4}{2B} + \sqrt{AB^3} \pi'^4 \cdot F_4, \\
a_9&=&\frac{\sqrt{A}}{J}\frac{\D}{\D r}\left(
J\sqrt{B}{\cal H}
\right), \\
b_1&=&\frac{1}{2}\sqrt{\frac{B}{A}}{\cal H},
\\
b_2&=&-2\sqrt{\frac{B}{A}}\Xi,
\\
b_3&=&\sqrt{\frac{B}{A}}\frac{1}{\pi'}
\left[
\left(2\pi''+\frac{B'}{B}\pi'\right)\Xi -2JJ'\left(\frac{A'}{A}-\frac{B'}{B}\right){\cal H}
+\frac{2J^2}{B}{\cal E}_A+4JJ''{\cal H}
\right]%=\frac{2}{A} \left( a_1'-a_2 \right)+\frac{2r^2}{\sqrt{AB} \phi'} {\cal E}_B
,
\\
b_4&=&\sqrt{\frac{B}{A}}\left(\pi'\Xi+2JJ'{\cal H}\right),
\\
b_5&=&-2b_1,\\
c_1&=&-\frac{1}{\sqrt{AB}}\Xi - 4\sqrt{\frac{ B^3}{A}} J J' \pi'^3 \cdot F_4,
\\
c_2&=&-\sqrt{AB} \left( \frac{A'}{2A} \Xi+JJ'\Gamma-\frac{J^2 \pi'}{X} \Sigma \right)
,
\\
c_3&=&J^2\sqrt{\frac{A}{B}} \frac{\partial {\cal E}_B}{\partial \pi},\\
c_4&=&\frac{1}{2}\sqrt{\frac{A}{B}} \Gamma + \sqrt{\frac{ B^3}{A}} \left(A' J + 2A J'\right) \pi'^3
 \cdot F_4,
\\
c_5&=&
-\frac{1}{2}\sqrt{AB}\left(\pi'\Gamma+\frac{A'}{A}{\cal H}
+\frac{2J'}{J}{\cal G}\right),
\\
c_6&=&\frac{J^2}{2} \sqrt{\frac{A}{B}} \left( \Sigma+\frac{A'B \pi'}{2J^2 A}\Xi+\frac{B\pi'J'}{J}\Gamma-\frac{1}{2} {\cal E}_B+\frac{BJ'^2}{J^2} {\cal G}+\frac{A'BJ'}{JA}{\cal H} \right),
\\
d_1&=&b_1,
\\
d_2&=&\sqrt{AB}\,\Gamma,
\\
d_3&=& \frac{\sqrt{AB}}{J^2}\left[
\frac{2JJ'}{\pi'}\left(\frac{A'}{A}-\frac{B'}{B}\right){\cal H}
-J^2\left(\frac{2J'}{J}-\frac{A'}{A}\right)\frac{\partial{\cal H}}{\partial\pi}
+\frac{2}{B\pi'}\left({\cal F}-{\cal G}\right)\right.
\nonumber\\&&\left.
-\frac{J^2}{2\pi'}\left(2\pi''+\frac{B'}{B}\pi'\right)
\left(\Gamma_1+\frac{2J'}{J}\Gamma_2\right)
-\frac{2J^2}{B\pi'}({\cal E}_A-{\cal E}_B)-\frac{4JJ''}{\pi'}{\cal H}
\right]
,
\\
d_4&=&\frac{\sqrt{AB}}{J^2}
\left({\cal G}-J^2{\cal E}_B\right),
\\
e_1&=&\frac{1}{2\sqrt{AB}}\left[
\frac{J^2}{X}({\cal E}_A-{\cal E}_B)-\frac{2}{\pi'}\Xi'+\left(\frac{A'}{A}
-\frac{X'}{X}\right)\frac{\Xi}{\pi'}
+\frac{2BJ'^2}{X}{\cal F}-\frac{2JJ'B}{X}{\cal H}' \right.\\&&\left.
-{\cal H}\frac{B^2J'^2}{JXA}\frac{\D}{\D r}\left(\frac{J^2A}{B}\right)+\frac{2BJJ''}{X}{\cal H}
\right] - \frac{4}{B \pi'^2} \cdot \frac{\mbox{d}}{\mbox{d} r}
\left[\sqrt{\frac{B^5}{A}} J J' \pi'^4 \cdot F_4\right], \nonumber \\
%&=&\frac{1}{A B \phi'} \left[ \left( \frac{A'}{A}+\frac{B'}{2B} \right) a_1 + a_2-2a_1' - 2rB a_6 \right], \\
e_2&=&-\sqrt{AB}\frac{J^2}{X}\Sigma,\\
e_3&=&J^2 \sqrt{\frac{A}{B}} \frac{\partial {\cal E}_\pi}{\partial \pi},
\end{eqnarray}
\begin{eqnarray}
% c_6&=&\frac{J^2}{2} \sqrt{\frac{A}{B}} \left( \Sigma+\frac{A'B \pi'}{2J^2 A}\Xi+\frac{B\pi'J'}{J}\Gamma-\frac{1}{2} {\cal E}_B+\frac{BJ'^2}{J^2} {\cal G}+\frac{A'BJ'}{JA}{\cal H} \right),
% \\
% d_1&=&b_1,
% \\
% d_2&=&\sqrt{AB}\,\Gamma,
% \\
% d_3&=& \frac{\sqrt{AB}}{J^2}\left[
% \frac{2JJ'}{\pi'}\left(\frac{A'}{A}-\frac{B'}{B}\right){\cal H}
% -J^2\left(\frac{2J'}{J}-\frac{A'}{A}\right)\frac{\partial{\cal H}}{\partial\pi}
% +\frac{2}{B\pi'}\left({\cal F}-{\cal G}\right)\right.
% \nonumber\\&&\left.
% -\frac{J^2}{2\pi'}\left(2\pi''+\frac{B'}{B}\pi'\right)
% \left(\Gamma_1+\frac{2J'}{J}\Gamma_2\right)
% -\frac{2J^2}{B\pi'}({\cal E}_A-{\cal E}_B)-\frac{4JJ''}{\pi'}{\cal H}
% \right]
% ,
% \\
% d_4&=&\frac{\sqrt{AB}}{J^2}
% \left({\cal G}-J^2{\cal E}_B\right), \\
% e_1&=&\frac{1}{2\sqrt{AB}}\left[
% \frac{J^2}{X}({\cal E}_A-{\cal E}_B)-\frac{2}{\pi'}\Xi'+\left(\frac{A'}{A}
% -\frac{X'}{X}\right)\frac{\Xi}{\pi'}
% +\frac{2BJ'^2}{X}{\cal F}-\frac{2JJ'B}{X}{\cal H}' \right.\\&&\left.
% -{\cal H}\frac{B^2J'^2}{JXA}\frac{\D}{\D r}\left(\frac{J^2A}{B}\right)+\frac{2BJJ''}{X}{\cal H}
% \right] - e_1^{BH}, \nonumber \\
% %&=&\frac{1}{A B \phi'} \left[ \left( \frac{A'}{A}+\frac{B'}{2B} \right) a_1 + a_2-2a_1' - 2rB a_6 \right], \\
% e_2&=&-\sqrt{AB}\frac{J^2}{X}\Sigma,\\
% e_3&=&J^2 \sqrt{\frac{A}{B}} \frac{\partial {\cal E}_\pi}{\partial \phi}, \\
e_4&=&\frac{\sqrt{AB}J'^2}{8X}\left(-\frac{4{\cal G}}{J^2}-\frac{4({\cal E}_A-{\cal E}_B)}{BJ'^2}+\frac{2A'{\cal H}'}{AJ'^2}+
\frac{4{\cal G}'}{JJ'}+\frac{4}{BJ^2J'^2}\left(1-\frac{JJ'BA'}{A}\right){\cal F}
\nonumber\right.\\ &&\left.-
\frac{4{\cal H}}{BJ^2J'^2}\left(1-BJ'^2(1+\frac{2A'J}{AJ'})+J(B'J'+2BJ'')\right)-\frac{2\pi'}{J'^2}\Gamma'
\nonumber\right.\\ &&\left.
+\frac{2\pi'}{JJ'}\left(-2+\frac{A'J}{AJ'}\right)\frac{\partial{\cal H}}{\partial\pi}
+\frac{\Xi\pi'}{J^3J'}\left(2-\frac{A'J}{AJ'}\right)\left[\frac{A'BJJ'}{A}-2+2BJ'^2\right]
\nonumber\right.\\ &&\left.
+\frac{\Gamma_1\pi'}{2J'}\left[\frac{4}{J}+\frac{A'^2BJ}{A^2}-\frac{4A'}{AJ'}-\frac{4BJ'^2}{J}
-\frac{2B'}{BJ'}+\frac{2BJ''}{BJ'^2}-\frac{4\pi''}{\pi'J'}
\right]
-\nonumber\right.\\ &&\left.
-
\frac{\Gamma_2\pi'}{J'}\left[\frac{2A'}{AJ}+\frac{A'^2}{A^2J'}(1-BJ'^2)-\frac{4J'}{J^2}(1-BJ'^2)+\frac{2B'}{BJ}+\frac{4\pi''}{\pi'J}
\right]\right) - e_4^{BH},
\\
% a_2^{BH}&=& 2  \sqrt{AB^3} \pi'^3 \cdot F_4 ,
% \\
% a_6^{BH}&=& - \sqrt{AB} \pi'^2 \left( B' \pi' + 2 B \pi''\right) \cdot F_4,
% \\
% a_8^{BH}&=& - \sqrt{AB^3} \pi'^4 \cdot F_4 ,
% \\
% c_1^{BH}&=& 4\sqrt{\frac{ B^3}{A}} J J' \pi'^3 \cdot F_4,
% \\
% c_4^{BH}&=& - \sqrt{\frac{ B^3}{A}} \left(A' J + 2A J'\right) \pi'^3
%  \cdot F_4,
% \\
% e_1^{BH}&=& \frac{4}{B \pi'^2} \cdot \frac{\mbox{d}}{\mbox{d} r}
% \left[\sqrt{\frac{B^5}{A}} J J' \pi'^4 \cdot F_4\right],
% \\
e_4^{BH}&=& -\frac{\pi'^2}{J^2 J'} \sqrt{\frac{B^3}{A}}
\left(A'' J^2 J' - 4 A J'^3 +A' J(J'^2 - J J'')\right) \cdot F_4
\\\nonumber&&
- \frac{A'J + 2AJ'}{B J^2 J' \pi'^2} \cdot
\frac{\mbox{d}}{\mbox{d} r} \left[\sqrt{\frac{B^5}{A}} J J' \pi'^4 \cdot F_4\right],
\\
a_{13}&=& J^2\; a_4,
\\
a_{14}&=& \sqrt{{A}{B}} \;J^2 \left[\cH' + \frac12 \left(\frac{B'}{B} +6\frac{J'}{J} \right)\cH \right],
\\
a_{15} &=& \sqrt{\frac{A}{B}}\; \cF,
\\
a_{16} &=& -\frac12 a_{15},
\\
%%%%%%%%%%%%%%%%%%%%%%%%%%%%
b_{8} &=& - 2\sqrt{\frac{B}{A}} J^2 \cH,
\\
b_{9} &=& \sqrt{\frac{B}{A}} J^2 \left(\frac{A'}{A} - 2 \frac{J'}{J}  \right) \cH,
\\
b_{10} &=& -b_6,
\\
b_{11} &=& \sqrt{\frac{B}{A}} \frac{A'}{A} \cH,
\\
c_{8} &=& -\frac{J^2}{\sqrt{{A}{B}}} (\cH - 2 F_4 B^2 \pi'^4),
\\
c_{9} &=& \frac14 \sqrt{{A}{B}} \left[\Gamma \pi' + \left(\frac{A'}{A} + 2 \frac{J'}{J} \right)\cH \right],
 \\
c_{10} &=& - \frac14\sqrt{{A}{B}} \left[
\frac{J'}{J} \left( \Gamma + \frac{J'}{J} \Gamma_2\right)\pi' - \frac14\frac{A'}{A}\left( \Gamma_1 - 2 \frac{J'}{J} \Gamma_2\right)\pi'+
\right.\\\nonumber&&\left.
\cH \frac{J'}{J} \left(\frac{A'}{A} + 3 \frac{J'}{J} \right)+
\frac{1}{BJ^2} (\cH - 2 F_4 B^2 \pi'^4) \right],
\\
c_{11} &=& - 
\frac12 \sqrt{{A}{B}} J^2  \left[\Gamma \pi' + \left(\frac{A'}{A} + 2 \frac{J'}{J} \right)\cH \right],
\\
c_{12} &=& - \frac12 \sqrt{{A}{B}} J^2 \left[
\frac{1}{2 J^2}\left(\frac{A'}{A} + \frac{J'}{J} \right)\Xi\pi' +
\frac34 \frac{J'}{J} \left( \Gamma_1 - 2 \frac{J'}{J} \Gamma_2\right)\pi' 
\right.\\\nonumber&&\left.
+ \frac{J'}{J} \left(\frac{A'}{A} + \frac{J'}{J} \right) \cH +
\frac{1}{BJ^2} (\cH - 2 F_4 B^2 \pi'^4) \right],
% \\
\end{eqnarray}
\begin{eqnarray}
%%%%%%%%%%%%%%%%%%%%%%%%%%%
c_{13} &=& \frac12 \frac{\sqrt{A}}{\sqrt{B}} (\cH - 2 F_4 B^2 \pi'^4),
\\
d_{5} &=& - \frac{\sqrt{{A}{B}}}{J^2} \cH,
\\
d_{6} &=& 2\frac{\sqrt{{A}{B}}J'}{J^3} \cH,
\\
d_{7} &=& a_4,
\\
p_{8} &=& -\frac{J^2}{2 \sqrt{{A}{B}}} \cF,
\\
p_{9} &=& \frac12 \sqrt{{A}{B}} J^2 \cH,
\end{eqnarray}

%%%%%%%%%%%%%%%%%%%%%%%%%%%%%%%%%%%%%%%%%%%%%%%%%%%%%%%%%%%%%%%%%%%%%%%%

% \section*{Appendix C}
% Coefficients $g_j$, $f_j$ and $h_j$ from action (48)

% % \begin{equation}
% \begin{eqnarray}
% g_1 &=&  \frac{B^{1/2} J}{2 (\ell^2+\ell-2)A^{3/2}} \cdot \frac{\cH}{\cF^2}
% \left[
% ( 2 A \cF' +  A' \cF) (2\cH' BJ +\cH B'J + 2 \cH B J') 
% \right.\\&& \left.
% - 2\cF A \frac{\D}{\D r} \left( 2\cH' BJ +\cH B'J + 2 \cH B J' \right)
% \right],
% \\
% g_2 &=&
% \end{eqnarray}

%%%%%%%%%%%%%%%%%%%%%%%%%%%%%%%%%%%%%%%%%%%%%%%%%%%%%%%%%%%%%%%%%%%%%%%%
\section*{Appendix C}
Here we put the rest of components of matrices $\bar{\mathcal{K}}$, $\bar{\mathcal{G}}$ and $\bar{\mathcal{M}}^{(\ell^2)}$ entering the quadratic
action~\eqref{eq:action_KGQM_new} for completeness:
\begin{equation}
\bar{\mathcal{K}}_{12} = \bar{\mathcal{K}}_{21} = \frac{B^{1/2}}{2(\ell+2)(\ell-1) A^{1/2}} \left[2(\ell+2)(\ell-1) (2\cH J J' + \Xi \pi') -
\frac{\left[\cH^2 B J^2 \right]'}{\cH B J} \frac{\TT}{\cF} \right],
\end{equation}
\begin{equation}
\bar{\mathcal{K}}_{22} =  \frac{4 \ell(\ell+1) A^{1/2} B^{1/2}}{\TT^2} \left[\ell(\ell+1) \frac{B}{A} (2 \mathcal{P}_1 -\cF) (2\cH J J' + \Xi \pi')^2
+ (\ell+2)(\ell-1) \cF\; \bar{\mathcal{K}}_{12}^2 \right],
\end{equation}

\begin{equation}
\bar{\mathcal{G}}_{12} = \bar{\mathcal{G}}_{21} = AB\frac{\cG}{\cF} \bar{\mathcal{K}}_{12}
\end{equation}

\begin{equation*}
\bar{\mathcal{G}}_{22} = \frac{4\ell(\ell+1) A^{1/2} B^{1/2}}{\TT^2} \frac{\cG}{\cH} \left[ \ell(\ell+1) B^2 (2 J^2 \Gamma \mathcal{H} \Xi \pi'^2  - \mathcal{G} \Xi^2 \pi'^2-4 J^4 \Sigma \mathcal{H}^2/B )
+ \frac{(\ell+2)(\ell-1)\cF^2}{AB \;\cG} \bar{\mathcal{G}}_{12}^2\right],
\end{equation*}

\begin{equation}
\bar{\mathcal{M}}^{(\ell^2)}_{12} =  \bar{\mathcal{M}}^{(\ell^2)}_{21} = - \ell(\ell+1)  \frac{ \cH (\cH - 2 F_4 B^2 \pi'^2) } {2 \cF J} \frac{A^{1/2}}{B^{1/2}} \left[ 2 \frac{\left[\cH^2 B J^2 \right]'}{\cH^2 J} + \frac{B(A'J - 2 A J')}{A} - \frac{B}{A} \cF \mathcal{P}_4
 \right],
\end{equation}

% \begin{equation*}
% \begin{aligned}
\begin{multline}
 \bar{\mathcal{M}}^{(\ell^2)}_{22} = -\ell(\ell+1) \;\frac{ \cH^2}{\cF} \frac{A^{1/2} B^{1/2}}{J^2} \left[ 2\frac{ \cF \cG }{\cH^2} + \frac{B^{1/2}}{A^{1/2}} J^4\; \cF 
\left[\frac{B^{1/2}}{A^{1/2}J^3} \mathcal{P}_4\right]' +\frac{1}{B} \left(\frac{\left[\cH^2 B J^2 \right]'}{\cH^2 J}\right)^2
 \right] \\
  - \frac{2}{(\cH - 2 F_4 B^2 \pi'^2)} \frac{\left[\cH^2 B J^2 \right]'}{\cH J^2} \bar{\mathcal{M}}_{12}^{(\ell^2)}.
%  \end{aligned}
% \end{equation*}
\end{multline}
%%%%%%%%%%%%%%%%%%%%%%%%%%%%%%%%%%%%%%%%%%%%%%%%%%%%%%%%%%%%%%%%%%%%%%%%

%%%%%%%%%%%%%%%%%%%%%%%%%%%%%%%%%%%%%%%%%%%%%%%%%%%%%%%%%%%%%%%%%%%%%%%%


\begin{thebibliography}{99}

\bibitem{Ellis} H.~G.~Ellis,
  ``Ether flow through a drainhole - a particle model in general relativity,''
  J.\ Math.\ Phys.\  {\bf 14} (1973) 104.
  %doi:10.1063/1.1666161
 
  \bibitem{Bronnikov}  K.~A.~Bronnikov,
  ``Scalar-tensor theory and scalar charge,''
  Acta Phys.\ Polon.\ B {\bf 4} (1973) 251.
  
\bibitem{Morris:1988cz}
  M.~S.~Morris and K.~S.~Thorne,
  ``Wormholes in space-time and their use for interstellar travel: A tool for teaching general relativity,''
  Am.\ J.\ Phys.\  {\bf 56} (1988) 395.{}
  %doi:10.1119/1.15620

\bibitem{Morris:1988tu}
  M.~S.~Morris, K.~S.~Thorne and U.~Yurtsever,
  ``Wormholes, Time Machines, and the Weak Energy Condition,''
  Phys.\ Rev.\ Lett.\  {\bf 61} (1988) 1446.
  %doi:10.1103/PhysRevLett.61.1446

  \bibitem{Visser}
M. Visser, "Lorentzian wormholes: From Einstein to Hawking," Woodbury, USA: AIP (1995) 412 p.

\bibitem{Hochberg_Visser}
D.~Hochberg and M.~Visser,
``General dynamic wormholes and violation of the null energy condition,''
[arXiv:gr-qc/9901020 [gr-qc]].

\bibitem{Dubovsky:2005xd}
S.~Dubovsky, T.~Gregoire, A.~Nicolis and R.~Rattazzi,
%``Null energy condition and superluminal propagation,''
JHEP \textbf{03}, 025 (2006)
% doi:10.1088/1126-6708/2006/03/025
[arXiv:hep-th/0512260 [hep-th]].

\bibitem{RubakovNEC}
  V.~A.~Rubakov,
  ``The Null Energy Condition and its violation,''
  Phys.\ Usp.\  {\bf 57} (2014) 128
   [Usp.\ Fiz.\ Nauk {\bf 184} (2014) no.2,  137]
  %doi:10.3367/UFNe.0184.201402b.0137
  % \href{https://arxiv.org/pdf/1401.4024.pdf}{[arXiv:1401.4024 [hep-th]]}.
  [arXiv:1401.4024 [hep-th]].
  %%CITATION = doi:10.3367/UFNe.0184.201402b.0137;%%

%These are from Maldacena
\bibitem{Gao:2016bin}
P.~Gao, D.~L.~Jafferis and A.~C.~Wall,
``Traversable Wormholes via a Double Trace Deformation,''
JHEP \textbf{12}, 151 (2017)
% doi:10.1007/JHEP12(2017)151
[arXiv:1608.05687 [hep-th]].

\bibitem{Fu:2019vco}
Z.~Fu, B.~Grado-White and D.~Marolf,
``Traversable Asymptotically Flat Wormholes with Short Transit Times,''
Class. Quant. Grav. \textbf{36}, no.24, 245018 (2019)
% doi:10.1088/1361-6382/ab56e4
[arXiv:1908.03273 [hep-th]].

\bibitem{Maldacena2018}
J.~Maldacena, A.~Milekhin and F.~Popov,
``Traversable wormholes in four dimensions,''
[arXiv:1807.04726 [hep-th]].



\bibitem{Bronnikov:2013coa}
K.~A.~Bronnikov, L.~N.~Lipatova, I.~D.~Novikov and A.~A.~Shatskiy,
``Example of a stable wormhole in general relativity,''
Grav. Cosmol. \textbf{19}, 269-274 (2013)
% doi:10.1134/S0202289313040038
[arXiv:1312.6929 [gr-qc]].


%phantom scalar field
\bibitem{Kodama}
   T.~Kodama,
  ``General Relativistic Nonlinear Field: A Kink Solution in a Generalized Geometry,''
  Phys.\ Rev.\ D {\bf 18} (1978) 3529.
  %doi:10.1103/PhysRevD.18.3529
  %%CITATION = doi:10.1103/PhysRevD.18.3529;%%

\bibitem{Armendariz}
  C.~Armendariz-Picon,
  ``On a class of stable, traversable Lorentzian wormholes in classical general relativity,''
  Phys.\ Rev.\ D {\bf 65} (2002) 104010
  %doi:10.1103/PhysRevD.65.104010
  % \href{https://arxiv.org/pdf/gr-qc/0201027.pdf}{[gr-qc/0201027]}.
  [gr-qc/0201027].

\bibitem{Gonzalez:2008wd}
J.~A.~Gonzalez, F.~S.~Guzman and O.~Sarbach,
``Instability of wormholes supported by a ghost scalar field. I. Linear stability analysis,''
Class. Quant. Grav. \textbf{26}, 015010 (2009)
% doi:10.1088/0264-9381/26/1/015010
[arXiv:0806.0608 [gr-qc]].



\bibitem{Tipler} 
F.~J.~Tipler,
  ``Energy conditions and spacetime singularities,''
  Phys.\ Rev.\ D {\bf 17} (1978) 2521.
  % doi:10.1103/PhysRevD.17.2521
 
\bibitem{Cline:2003gs} 
  J.~M.~Cline, S.~Jeon and G.~D.~Moore,
  ``The Phantom menaced: Constraints on low-energy effective ghosts,''
  Phys.\ Rev.\ D {\bf 70}, 043543 (2004)
  % \href{https://arxiv.org/pdf/hep-ph/0311312.pdf}{[hep-ph/0311312]}.
[hep-ph/0311312].

%%Wormholes in modified gravity

\bibitem{Lobo}
F.~S.~N.~Lobo and M.~A.~Oliveira,
``Wormhole geometries in f(R) modified theories of gravity,''
Phys. Rev. D \textbf{80}, 104012 (2009)
% doi:10.1103/PhysRevD.80.104012
[arXiv:0909.5539 [gr-qc]].

\bibitem{Agnese}
A.~G.~Agnese and M.~La Camera,
``Wormholes in the Brans-Dicke theory of gravitation,''
Phys. Rev. D \textbf{51}, 2011-2013 (1995)
% doi:10.1103/PhysRevD.51.2011

\bibitem{Nandi}
K.~K.~Nandi, B.~Bhattacharjee, S.~M.~K.~Alam and J.~Evans,
``Brans-Dicke wormholes in the Jordan and Einstein frames,''
Phys. Rev. D \textbf{57}, 823-828 (1998)
% doi:10.1103/PhysRevD.57.823
[arXiv:0906.0181 [gr-qc]].

\bibitem{Kanti}
P.~Kanti, B.~Kleihaus and J.~Kunz,
``Stable Lorentzian Wormholes in Dilatonic Einstein-Gauss-Bonnet Theory,''
Phys. Rev. D \textbf{85}, 044007 (2012)
% doi:10.1103/PhysRevD.85.044007
[arXiv:1111.4049 [hep-th]].

%%% STT theory
\bibitem{Horndeski:1974wa}
  G.~W.~Horndeski,
  ``Second-order scalar-tensor field equations in a four-dimensional space,''
  Int.\ J.\ Theor.\ Phys.\  {\bf 10} (1974) 363.
  %doi:10.1007/BF01807638

\bibitem{Gleyzes:2014dya}
  J.~Gleyzes, D.~Langlois, F.~Piazza and F.~Vernizzi,
  ``Healthy theories beyond Horndeski,''
  Phys.\ Rev.\ Lett.\  {\bf 114} (2015) no.21,  211101;
  %doi:10.1103/PhysRevLett.114.211101
  % \href{https://arxiv.org/pdf/1404.6495.pdf}{[arXiv:1404.6495 [hep-th]]}.
  [arXiv:1404.6495 [hep-th]].

\bibitem{Gleyzes:2014qga}
  J.~Gleyzes, D.~Langlois, F.~Piazza and F.~Vernizzi,
  ``Exploring gravitational theories beyond Horndeski,''
  JCAP {\bf 1502} (2015) 018
  %doi:10.1088/1475-7516/2015/02/018
  % \href{https://arxiv.org/pdf/1408.1952.pdf}{[arXiv:1408.1952 [astro-ph.CO]]}.
  [arXiv:1408.1952 [astro-ph.CO]].

\bibitem{Langlois1}
D.~Langlois and K.~Noui,
``Degenerate higher derivative theories beyond Horndeski: evading the Ostrogradski instability,''
JCAP \textbf{02}, 034 (2016)
% doi:10.1088/1475-7516/2016/02/034
[arXiv:1510.06930 [gr-qc]].

\bibitem{Langlois2}
D.~Langlois,
``Dark energy and modified gravity in degenerate higher-order scalar\textendash{}tensor (DHOST) theories: A review,''
Int. J. Mod. Phys. D \textbf{28}, no.05, 1942006 (2019)
% doi:10.1142/S0218271819420069
[arXiv:1811.06271 [gr-qc]].

\bibitem{KobaRev}
T.~Kobayashi,
``Horndeski theory and beyond: a review,''
Rept. Prog. Phys. \textbf{82}, no.8, 086901 (2019)
% doi:10.1088/1361-6633/ab2429
[arXiv:1901.07183 [gr-qc]].


%%Horndeski and beyond Horndeski

\bibitem{Bronnikov:2010tt}
  K.~A.~Bronnikov, M.~V.~Skvortsova and A.~A.~Starobinsky,
  ``Notes on wormhole existence in scalar-tensor and F(R) gravity,''
  Grav.\ Cosmol.\  {\bf 16} (2010) 216
  %doi:10.1134/S0202289310030047
  % \href{https://arxiv.org/pdf/1005.3262.pdf}{[arXiv:1005.3262 [gr-qc]]}.
[arXiv:1005.3262 [gr-qc]].

\bibitem{Korolev}
R.~V.~Korolev and S.~V.~Sushkov,
``Exact wormhole solutions with nonminimal kinetic coupling,''
Phys. Rev. D \textbf{90}, 124025 (2014)
% doi:10.1103/PhysRevD.90.124025
[arXiv:1408.1235 [gr-qc]].

\bibitem{Rubakov:2015gza}
  V.~A.~Rubakov,
  ``Can Galileons support Lorentzian wormholes?,''
  Teor.\ Mat.\ Fiz.\  {\bf 187} (2016) no.2,  338
   [Theor.\ Math.\ Phys.\  {\bf 187} (2016) no.2,  743]
  %doi:10.1134/S004057791605010X
   % \href{https://arxiv.org/pdf/1509.08808.pdf}{[arXiv:1509.08808 [hep-th]]}.
   [arXiv:1509.08808 [hep-th]].

\bibitem{Rubakov:2016zah}
  V.~A.~Rubakov,
  ``More about wormholes in generalized Galileon theories,''
  Theor.\ Math.\ Phys.\  {\bf 188} (2016) no.2,  1253
   [Teor.\ Mat.\ Fiz.\  {\bf 188} (2016) no.2,  337]
  %doi:10.1134/S0040577916080080
  % \href{https://arxiv.org/pdf/1601.06566.pdf}{[arXiv:1601.06566 [hep-th]]}.
  [arXiv:1601.06566 [hep-th]].

\bibitem{Kolevatov:2016ppi}
  R.~Kolevatov and S.~Mironov,
  ``Cosmological bounces and Lorentzian wormholes in Galileon theories with an extra scalar field,''
  Phys.\ Rev.\ D {\bf 94} (2016) no.12,  123516
  %doi:10.1103/PhysRevD.94.123516
   % \href{https://arxiv.org/pdf/1607.04099.pdf}{[arXiv:1607.04099 [hep-th]]}.
   [arXiv:1607.04099 [hep-th]].

\bibitem{Trincherini}
G.~Franciolini, L.~Hui, R.~Penco, L.~Santoni and E.~Trincherini,
``Stable wormholes in scalar-tensor theories,''
JHEP \textbf{01}, 221 (2019)
% doi:10.1007/JHEP01(2019)221
[arXiv:1811.05481 [hep-th]].

\bibitem{wormhole1}  
S.~Mironov, V.~Rubakov and V.~Volkova,
``More about stable wormholes in beyond Horndeski theory,''
Class. Quant. Grav. \textbf{36}, no.13, 135008 (2019)
% doi:10.1088/1361-6382/ab2574
[arXiv:1812.07022 [hep-th]].

\bibitem{KorolevLobo}
R.~Korolev, F.~S.~N.~Lobo and S.~V.~Sushkov,
``General constraints on Horndeski wormhole throats,''
Phys. Rev. D \textbf{101}, no.12, 124057 (2020)
% doi:10.1103/PhysRevD.101.124057
[arXiv:2004.12382 [gr-qc]].

\bibitem{Bakopoulos}
A.~Bakopoulos, C.~Charmousis and P.~Kanti,
``Traversable wormholes in beyond Horndeski theories,''
JCAP \textbf{05}, no.05, 022 (2022)
% doi:10.1088/1475-7516/2022/05/022
[arXiv:2111.09857 [gr-qc]].

%stability issues
\bibitem{Olegi}
  O.~A.~Evseev and O.~I.~Melichev,
  ``No static spherically symmetric wormholes in Horndeski theory,''
  Phys.\ Rev.\ D {\bf 97} (2018) no.12,  124040
  %doi:10.1103/PhysRevD.97.124040
  % \href{https://arxiv.org/pdf/1711.04152.pdf}{[arXiv:1711.04152 [gr-qc]]}.
  [arXiv:1711.04152 [gr-qc]].


%\cite{Kobayashi:2012kh}
\bibitem{Kobayashi:odd}
  T.~Kobayashi, H.~Motohashi and T.~Suyama,
  ``Black hole perturbation in the most general scalar-tensor theory with second-order field equations I: the odd-parity sector,''
  Phys.\ Rev.\ D {\bf 85} (2012) 084025
   Erratum: [Phys.\ Rev.\ D {\bf 96} (2017) no.10,  109903]
  %doi:10.1103/PhysRevD.96.109903, 10.1103/PhysRevD.85.084025
  % \href{https://arxiv.org/pdf/1202.4893.pdf}{ [arXiv:1202.4893 [gr-qc]]}.
  [arXiv:1202.4893 [gr-qc]].

%\cite{Kobayashi:2014wsa}
\bibitem{Kobayashi:even}
  T.~Kobayashi, H.~Motohashi and T.~Suyama,
  ``Black hole perturbation in the most general scalar-tensor theory with second-order field equations II: the even-parity sector,''
  Phys.\ Rev.\ D {\bf 89} (2014) no.8,  084042
  %doi:10.1103/PhysRevD.89.084042
 % \href{https://arxiv.org/pdf/1402.6740.pdf}{[arXiv:1402.6740 [gr-qc]]}.
[arXiv:1402.6740 [gr-qc]].

\bibitem{Ijjas}
A.~Ijjas,
``Space-time slicing in Horndeski theories and its implications for non-singular bouncing solutions,''
JCAP \textbf{02}, 007 (2018)
% doi:10.1088/1475-7516/2018/02/007
[arXiv:1710.05990 [gr-qc]].

\bibitem{GammaCrossing}
S.~Mironov, V.~Rubakov and V.~Volkova,
``Bounce beyond Horndeski with GR asymptotics and $\gamma$-crossing,''
JCAP \textbf{10}, 050 (2018)
% doi:10.1088/1475-7516/2018/10/050
[arXiv:1807.08361 [hep-th]].

%main text

\bibitem{ReggeWheeler}
  T.~Regge and J.~A.~Wheeler,
  ``Stability of a Schwarzschild singularity,''
  Phys.\ Rev.\  {\bf 108} (1957) 1063.
  %doi:10.1103/PhysRev.108.1063

   \bibitem{Gerlach:1979rw}
  U.~H.~Gerlach and U.~K.~Sengupta,
  ``Gauge Invariant Perturbations On Most General Spherically Symmetric Space-times,''
  Phys.\ Rev.\ D {\bf 19} (1979) 2268.
  %doi:10.1103/PhysRevD.19.2268
  %%CITATION = doi:10.1103/PhysRevD.19.2268;%%

\end{thebibliography}
\end{document}